\documentclass[twocolumn,tighten,twocolappendix]{aastex63mod}
\usepackage{amsmath}  
\usepackage{bm}  

\newcommand*{\mt}{\mathrm}
\newcommand*{\abt}{\mathord{\sim}} 
\newcommand*{\dtl}{\mathrm{d}}
\renewcommand{\vec}[1]{\boldsymbol{#1}}  
\newcommand*{\bp}{\beta_{\mathrm{p}}}
\newcommand*{\comp}{{c/\omega_\mathrm{pe}}}
\newcommand*{\Ma}{\mathcal{M}_{\mathrm{A}}}
\newcommand*{\Mms}{\mathcal{M}_{\mathrm{ms}}}
\newcommand*{\Ms}{\mathcal{M}_{\mathrm{s}}}
\newcommand*{\me}{m_{\mathrm{e}}}
\newcommand*{\mi}{m_{\mathrm{i}}}
\newcommand*{\mime}{m_{\mathrm{i}}/m_{\mathrm{e}}}
\newcommand*{\Omce}{\Omega_{\mathrm{e}}}
\newcommand*{\Omci}{\Omega_{\mathrm{i}}}
\newcommand*{\omp}{\omega_{\mathrm{pe}}}
\newcommand*{\ompi}{\omega_{\mathrm{pi}}}
\newcommand*{\rLi}{r_{\mathrm{Li}}}
\newcommand*{\Te}{T_{\mathrm{e}}}
\newcommand*{\TeTi}{T_{\mathrm{e}}/T_{\mathrm{i}}}
\newcommand*{\Ti}{T_{\mathrm{i}}}
\newcommand*{\kB}{k_{\mathrm{B}}}

\shorttitle{Perpendicular shock electron heating (draft, \today)}  
\shortauthors{Tran and Sironi (draft, \today)}  

\begin{document}

\title{Electron Heating in Perpendicular Low-Beta Shocks}

\correspondingauthor{Aaron Tran}
\email{aaron.tran@columbia.edu}

\author[0000-0003-3483-4890]{Aaron Tran}
\affiliation{Department of Astronomy, Columbia University \\
550 W 120th St.~MC~5246, New York, NY 10027, USA}

\author[0000-0002-5951-0756]{Lorenzo Sironi}
\affiliation{Department of Astronomy, Columbia University \\
550 W 120th St.~MC~5246, New York, NY 10027, USA}

\begin{abstract}
Collisionless shocks heat electrons in the solar wind, interstellar blast
waves, and hot gas permeating galaxy clusters.
How much shock heating goes to electrons instead of ions,
and what plasma physics controls electron heating?
We simulate 2-D perpendicular shocks with a fully kinetic particle-in-cell code.
For magnetosonic Mach number
$\mathcal{M}_\mathrm{ms} \sim 1$--$10$ and plasma beta $\beta_\mathrm{p}
\lesssim 4$, the post-shock electron/ion temperature ratio
$T_\mathrm{e}/T_\mathrm{i}$ decreases from $1$ to $0.1$ with increasing
$\mathcal{M}_\mathrm{ms}$.
In a representative $\mathcal{M}_\mathrm{ms}=3.1$, $\beta_\mathrm{p}=0.25$
shock, electrons heat above adiabatic compression in two steps:
ion-scale $E_\parallel = \vec{E} \cdot \hat{\vec{b}}$ accelerates electrons into
streams along $\vec{B}$, which then relax via two-stream-like instability.
The $\vec{B}$-parallel heating is mostly induced by waves;
$\vec{B}$-perpendicular heating is mostly adiabatic compression by
quasi-static fields.
\end{abstract}

\keywords{Plasma astrophysics (1261), Space plasmas (1544), Planetary bow shocks (1246), Shocks (2086)}

\section{Introduction}

Electron heating in collisionless shocks -- stated as post-shock electron/ion
temperature ratio $\TeTi$ -- is not constrained by magnetohydrodynamic (MHD)
shock jump conditions.  How much do electrons heat, and how do they heat?
A prediction for $\TeTi$ can constrain models for gas accretion onto galaxy
clusters \citep{avestruz2015}, and cosmic ray acceleration in supernova remnants
\citep{helder2010, yamaguchi2014, hovey2018}.
Detailed study of the electron heating physics can also help us interpret new
high-resolution data from the Magnetospheric Multiscale Mission \citep{chen2018,
goodrich2018, cohen2019}.

In the heliosphere, shocks of magnetosonic Mach number $\Mms \gtrsim 2$--$3$
heat electrons beyond adiabatic compression via a two-step process:
electrons accelerate in bulk along $\vec{B}$ towards the shock downstream, then
relax into ``flat-top'' distributions in $\vec{B}$-parallel velocity
\citep{feldman1982, feldman1983-earth, chen2018}.
Two mechanisms -- quasi-static direct current (DC) fields and plasma waves --
may drive $\vec{B}$-parallel acceleration.
In the DC mechanism, an electric potential jump in the shock layer (i.e., a
quasi-static electric field that points along shock normal) accelerates
electrons in bulk
\citep{feldman1983-earth, goodrich1984, scudder1986-iii, scudder1996, hull2001,
lefebvre2007, schwartz2014}.
The DC electron energy gain scales with $\cos^2\theta$, where $\theta$ is the
angle between $\vec{B}$ and shock
normal \citep{goodrich1984}.  We expect no heating in exactly planar
perpendicular shocks, but shock rippling from ion-scale waves
\citep{lowe2003, johlander2016, hanson2019} can bend $\vec{B}$, alter
$\theta$, and enable DC heating.
Plasma waves with non-zero $E_\parallel = \vec{E} \cdot \hat{\vec{b}}$, such as
oblique whistlers, can
also provide electron bulk acceleration and thus heating \citep{wilson2014-i,
wilson2014-ii}.
Such plasma waves are intrinsic to shock structure \citep{wilson2009, wilson2012,
krasnoselskikh2002, dimmock2019} and may be sustained by free energy from,
e.g., shock-reflected ions \citep{wu1984, matsukiyo2006, muschietti2017}.

In this Letter, we study thermal electron heating in multi-dimensional
particle-in-cell (PIC) simulations of perpendicular shocks with realistic
structure
(requiring high ion/electron mass ratio $\mime$
\citep{krauss-varban1995, umeda2012-mime, umeda2014})
and high grid resolution to resolve
electron scattering and relaxation after $\vec{B}$-parallel bulk acceleration.

\section{Method}

We simulate collisionless 2-D ($x$-$y$) ion-electron shocks using the
relativistic particle-in-cell (PIC) code TRISTAN-MP
\citep{buneman1993, spitkovsky2005}.
We inject plasma with velocity $-u_0 \hat{\vec{x}}$ and magnetic field $B_0 \hat{\vec{y}}$
from the simulation domain's right-side (upstream) boundary.
Injected plasma reflects from a conducting wall at $x=0$, forming a shock that
travels towards $+\hat{\vec{x}}$.
The shocked downstream plasma has zero bulk velocity, and the upstream
$\vec{B}$ is perpendicular to the shock normal, so
$\theta = 90^{\circ}$.
The simulation domain expands along $+\hat{\vec{x}}$ to keep the right-side boundary
$\gtrsim 1.5\;\rLi$ ahead of the shock front \cite[Sec.~2]{sironi2009}, where
$\rLi = u_0 / \Omci$ is a characteristic ion Larmor radius;
we checked that shock heating physics is not artifically affected by the
right-side boundary.
Upstream ions and electrons have equal density $n_0$ and temperature $T_0$.
The plasma frequencies
$\omega_{\mathrm{p\{i,e\}}} = \sqrt{4 \pi n_0 e^2/m_\mathrm{\{i,e\}}}$
and cyclotron frequencies
$\Omega_\mathrm{\{i,e\}} = e B_0 / (m_\mathrm{\{i,e\}} c)$
where subscripts $\mathrm{i}$ and $\mathrm{e}$ denote ions and electrons.
We use Gaussian CGS units throughout.

Our fiducial simulations have ion/electron mass ratio $\mime=625$
and total plasma beta $\bp = 16 \pi n_0 \kB T_0 / {B_0}^2 = 0.25$.
The fast magnetosonic, sonic, and Alfv\'{e}n Mach numbers are
$\Mms = u_\mt{sh} / \sqrt{{c_\mt{s}}^2 + {v_\mt{A}}^2} = 1$--$10$,
$\Ms = u_\mt{sh} / c_\mt{s} = 3$--$20$,
and $\Ma = u_\mt{sh} / v_\mt{A} = 1.5$--$10$.
The sound speed $c_\mt{s} = \sqrt{2 \Gamma \kB T_0/(m_i+m_e)}$,
Alfv\'{e}n speed $v_\mt{A} = {B_0} / \sqrt{4 \pi n_0 (m_i+m_e)}$,
and, $u_{\mathrm{sh}}$ is the speed of upstream plasma in the shock's rest
frame; for non-relativistic speeds, $u_\mathrm{sh} = u_0 / (1 - 1/r)$ where $r
\leq 4$ is the MHD shock-compression ratio.
The one-fluid adiabatic index $\Gamma$ is not known a priori, but it is set
self-consistently by the degree of ion and electron isotropization.
We report Mach numbers assuming $\Gamma=2$, which overestimates $\Mms$ by
$\abt1$--$10\%$ for stronger shocks that isotropize ions/electrons and have
$\Gamma \approx 5/3$.

\begin{figure}
    \begin{interactive}{animation}{{mov_shockstruct_lo-res}.mp4}
    \includegraphics[width=3.375in]{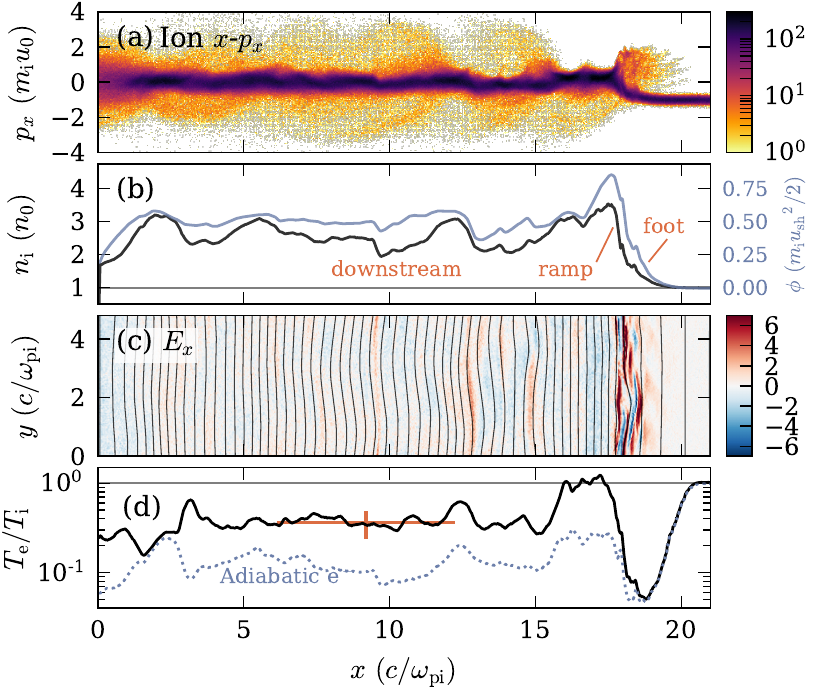}
    \end{interactive}
    \caption{
    Overview of $\Mms=3.1$, $\bp=0.25$ shock
    at $t=1253\;{\ompi}^{-1}=14.2\;{\Omci}^{-1}$.
    (a) Ion $x$-$p_x$ phase-space distribution for full domain, where $p_x$ is
    ion $x$-momentum normalized to upstream momentum $\mi u_0$.
    (b) Ion density $n_\mt{i}/n_0$ (black) and electromotive force
    $\phi(x) = -\int_\infty^x E_x(x') \dtl x'$ in units of $\mi {u_\mt{sh}}^2 /2$
    (light blue).
    Both curves are 1-D volume-weighted averages over $y$.
    The shock foot, ramp, and downstream are annotated.
    (c) Electric field $E_x$ normalized to upstream motional field $u_0 B_0/c$
    with magnetic field lines overlaid.
    $\vec{B}$ points up; i.e., along $+\hat{\vec{y}}$.
    (d) $\TeTi$, density-weighted $y$ average (black), compared to prediction
    for fluid adiabatic electron heating defined in text (blue dotted).
    Orange cross is starred measurement in Fig.~\ref{fig:machscale}(a); cross
    width is measurement region, and cross height is standard deviation over
    2-D region delimited by cross width.
    An animation of this figure in the online journal plots shows time
    evolution from $t=0$ to $1253\;{\ompi}^{-1}$ and demonstrates that the
    temperature ratio $\TeTi$ stabilizes $\abt 3 \;c/{\ompi}$ downstream
    of the shock ramp.
    }
    \label{fig:structure}
\end{figure}

The grid cell size $\Delta x = \Delta y = 0.1 \comp$ and the timestep
$\Delta t = 0.045 {\omp}^{-1}$ so that $c = 0.45 \Delta x / \Delta t$.
Upstream plasma has $16$ particles per cell per species.
We smooth the electric current with 32 sweeps of a three-point binomial
(``1-2-1'') filter at each timestep \citep[Appendix C]{birdsall1991}.
The $N=32$ sweeps approximate a Gaussian filter with standard deviation
$\sqrt{N/2} = 4$ cells or $0.4 c/\omp$.  The filter's half-power cut-off is at wavenumber
$k \approx \sqrt{2/N} (\Delta x)^{-1} = 2.5 \left(c/\omp\right)^{-1}$,
which implies 50\% damping at wavelengths
$\lambda \approx \pi \sqrt{2 N} \Delta = 2.5 \left(c/\omp\right)$.
Electron-scale waves may be damped, but we will later show that
electron-scale waves mainly scatter rather than heat.
We simulated 2-D $\Mms=3.1$, $\bp=0.25$ shocks with $4\times$ larger or
smaller sweep number $N$; the ratio $\TeTi$ did not change much.
We adjust $T_0$, $u_0$, and $B_0$ to control $\Mms$ and $\bp$ while keeping
shocked electrons non-relativistic;
i.e., post-shock $\kB T_\mt{e} \lesssim 0.05 m_\mt{e} c^2$.
The ratio $\tau = \omp/\Omce = 2.5$--$11$ ($\tau \gg 1$ for solar wind and
astrophysical settings).
The transverse ($y$) width is
$2.9$--$5.8\;c/\ompi = 720$--$1440$ cells.
Simulation durations are
$10$--$20 \;{\Omci}^{-1} = 932$--$2736\;{\ompi}^{-1}$
so that post-shock $\TeTi$ reaches steady state.
The temperatures $T_\mt{\{i,e\}}$, $T_{\mt{\{i,e\}}\parallel}$, and
$T_{\mt{\{i,e\}}\perp}$ are moments of the particle distribution in a $5^N$
cell region, where $N \in \{1,2,3\}$ is the domain dimensionality.
The co-moving frame boost for moment calculation uses a fluid velocity also
averaged over $5^N$ cells.
All $\parallel$- and $\perp$-subscripted quantities are taken with respect to
local $\vec{B}$.

Fig.~\ref{fig:structure} shows a representative simulation.
Ions transmit or reflect at the shock ramp (Fig.~\ref{fig:structure}(a-b)).
The shock front is rippled (Fig.~\ref{fig:structure}(c)).
A net $E_x$ potential exists across the shock, and $E_x$ is also modulated
by the $\vec{B}$ rippling wavelength (Fig.~\ref{fig:structure}(b-c)).
Reflected ions accelerate in the motional field $E_z = u_0 B_0 /c$ before
re-entering the shock \citep{leroy1982}; this lowers $\TeTi$ in the shock foot
(Fig.~\ref{fig:structure}(d)).
Electrons heat above adiabatic expectation in the shock ramp and settle to
$\TeTi \approx 0.4$; no appreciable heating occurs after the shock ramp
(Fig.~\ref{fig:structure}(d)).
The fluid adiabatic prediction in Fig.~\ref{fig:structure}(d) is
$T_\mathrm{e,ad} / (\Ti + \Te - T_\mathrm{e,ad})$,
using measured $\Ti$ and $\Te$ and assuming
$T_\mathrm{e,ad} = T_0 [1 + 2 (n/n_0)^{\Gamma-1}]/3$ with $\Gamma = 2$.

\begin{figure}
    \includegraphics[width=3.375in]{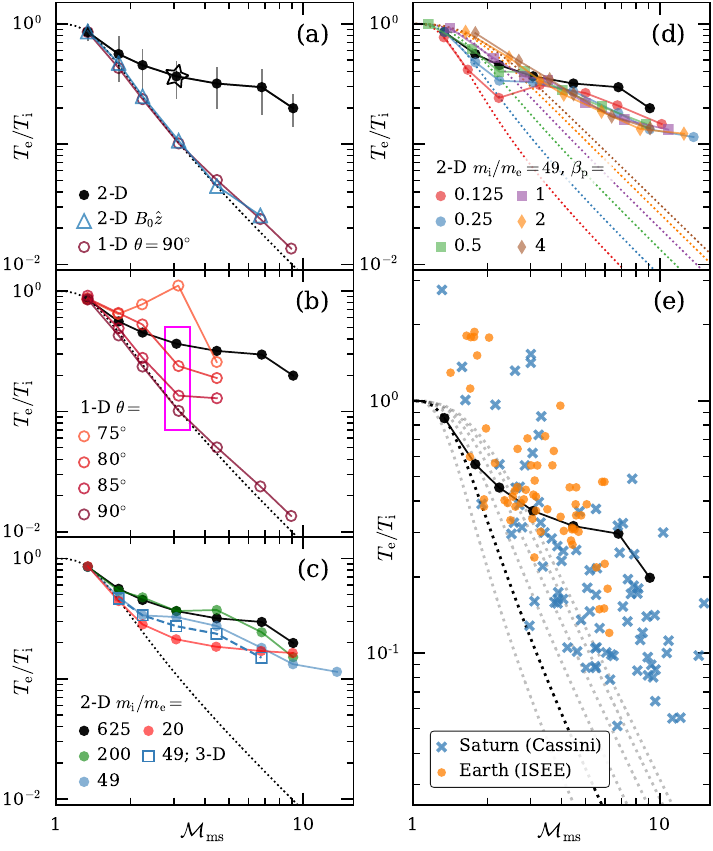}
    \caption{
    Post-shock $\TeTi$ dependence on $\Mms$ for various setups.
    All panels: black curve is $\TeTi$ from fiducial 2-D $\mime=625$,
    $\bp=0.25$ shocks.
    Dotted curves are expected $\TeTi$ from MHD shock jump conditions, assuming
    adiabatic electron heating alone.
    (a): Varying geometry.  2-D domain with out-of-plane $B_0 \hat{\vec{z}}$
    (triangles) and 1-D domain (hollow circles).
    Starred datum appears in
    Figs.~\ref{fig:structure},~\ref{fig:distrevol}--\ref{fig:edotv}.
    Error bars on black curve are standard deviation of $\TeTi$ within
    measurement region.
    (b): 1-D domain, varying $\theta$.
    Darkest hollow circles ($\theta=90^{\circ}$) same as (a).
    (c): 2-D domain, varying $\mime$.  Dashed cyan curve with square
    markers comprises 3-D $\mime=49$ simulations.
    (d): 2-D domain, $\mime=49$, varying $\bp$.  Dotted curves are adiabatic
    expectation as in panels (a-c), with $\bp$ increasing from left to right.
    (e): Comparison to solar wind bow shock measurements at Earth (orange
    circles) \citep{schwartz1988} and Saturn (blue crosses) \citep{masters2011}
    as compiled by \citep{ghavamian2013}.  Five Saturn measurements with
    $\Mms>20$ are not shown.
    }
    \label{fig:machscale}
\end{figure}

\section{Shock Parameter Scaling}

We measure post-shock $\TeTi$ (Fig.~\ref{fig:structure}(c))
as a function of $\Mms$ for many simulations with varying
dimensionality, magnetic field orientation $\theta$, $\mime$, and $\bp$.
We also adjust domain width, particle resolution, and current smoothing to
control noise and computing cost.
In simulations with $\theta < 90^{\circ}$, the
right-side boundary expands at $\max\left(u_\mathrm{sh}, 0.5c\cos\theta\right)$
to retain shock-reflected electrons streaming along $\vec{B}$.

We show the post-shock $\TeTi$ for our fiducial 2-D $\mime=625$ shocks with
in-plane upstream magnetic field $B_0 \hat{\vec{y}}$ in Fig.~\ref{fig:machscale}(a).
These fiducial simulations are converged in $\TeTi$ with respect to transverse
($y$) width.
For perpendicular shocks, we find that electron heating beyond adiabatic
compression requires 2-D geometry with in-plane $\vec{B}$.
Corresponding 2-D simulations with out-of-plane $\vec{B}$ (along $\hat{\vec{z}}$) and 1-D
simulations heat electrons by compression alone (Fig.~\ref{fig:machscale}(a)).
At $\Mms \sim 5$--$10$, the 2-D simulations with out-of-plane $\vec{B}$ and 1-D
simulations show weak super-adiabatic heating in the shock layer, but the
$\TeTi$ measurement is also less precise due to numerical heating.
\citet{shimada2000, shimada2005} saw strong electron heating in 1-D
perpendicular $\mime=20$ shocks due to Buneman instability between
shock-reflected ions and incoming electrons.  The higher $\mime=625$
suppresses the Buneman instability in our 1-D shocks.

Can DC heating in a 2-D rippled shock -- i.e., varying
local magnetic field angle due
to self-generated waves -- explain the super-adiabatic electron heating seen in
our fiducial 2-D simulations?
To estimate the DC heating from varying $\theta$, we perform 1-D oblique
shock simulations with varying $\theta < 90^{\circ}$
(Fig.~\ref{fig:machscale}(b)); recall that $\theta$ is the angle between
$\vec{B}$ and shock normal.
The 1-D setup keeps quasi-static shock structure (averaged over shock
reformation cycles)
and should retain DC heating while excluding waves oblique to the shock
normal.
We do find super-adiabatic heating in 1-D oblique shocks.
Electrons heat more for lower $\theta$, which is qualitatively consistent with
DC field heating \citep{goodrich1984}.
For our representative $\Mms = 3.1$ shock, which has local ripple
$\theta \gtrsim 80^{\circ}$ (Fig.~\ref{fig:edotv}(f)), the DC heating
inferred from 1-D oblique shock simulations appears too low to explain the full
amount of super-adiabatic heating (Fig.~\ref{fig:machscale}(b, box)).

Our fiducial 2-D perpendicular shocks appear converged in mass ratio at $\mime
\sim 200$--$625$ (Fig.~\ref{fig:machscale}(c)), consistent with prior simulations
\citep{umeda2012-mime, umeda2014}
and theory \citep{krauss-varban1995}.
For $\mime=20$--$625$, 2-D shocks agree on $\TeTi$ to within a factor of
$2$--$3$.
A set of 3-D $\mime=49$ simulations with narrower transverse width, $2.7
c/\ompi$, shows good agreement too.
Agreement between 2-D and 3-D for $\mime=49$ suggests that 2-D simulations with
in-plane $\vec{B}$ include the essential physics for electron heating.

To see how heating depends on $\bp$, we reduce $\mime$ to $49$ and sweep
$\bp$ over $0.125$--$4$ (Fig.~\ref{fig:machscale}(d)).
Electron heating increases above adiabatic at $\Mms \sim 2$--$3$ for all $\bp$.
At $\Mms \sim 3$--$5$ and $\bp \leq 1$, two-step $\vec{B}$-parallel electron
heating (which we describe below) operates for all $\bp \lesssim 1$.
At $\Mms \sim 3$--$5$ and $\bp \gtrsim 2$,
a distinct electron cyclotron whistler instability
is expected to heat electrons instead \citep{guo2017, guo2018}.
At $\Mms \gtrsim 5$, shock structure is more complex, which we do not explore
here.
The relationship between $\TeTi$ and $\Mms$ does not appear to depend on $\bp$
for $\Mms \gtrsim 4$.

Our fiducial $\TeTi$--$\Mms$ data are order-of-magnitude consistent with
measurements from solar wind bow shocks (Fig.~\ref{fig:machscale}(e)),
replotted from \citet{ghavamian2013}.
The Saturn data are uncertain in both $\TeTi$ and $\Mms$ due to a lack of ion
temperature measurements from Cassini \citep{masters2011}, so $\Mms = 0.671 \Ma$
(equivalent to $\bp \sim 1.5$) is assumed following
\citet{ghavamian2013}; we note $\bp \sim 1.5$ is a typical value
\citep{richardson2002}.
The Earth data have $\bp \sim 0.1$--$1$ and use directly measured ion and
electron temperatures from the ISEE spacecraft \citep{schwartz1988}.
Both datasets are mostly quasi-perpendicular, with a majority of shocks having
$50^\circ < \theta < 90^\circ$ \citep{schwartz1988, masters2011}.

\section{Electron Heating Physics}

\begin{figure}
    \includegraphics[width=3.375in]{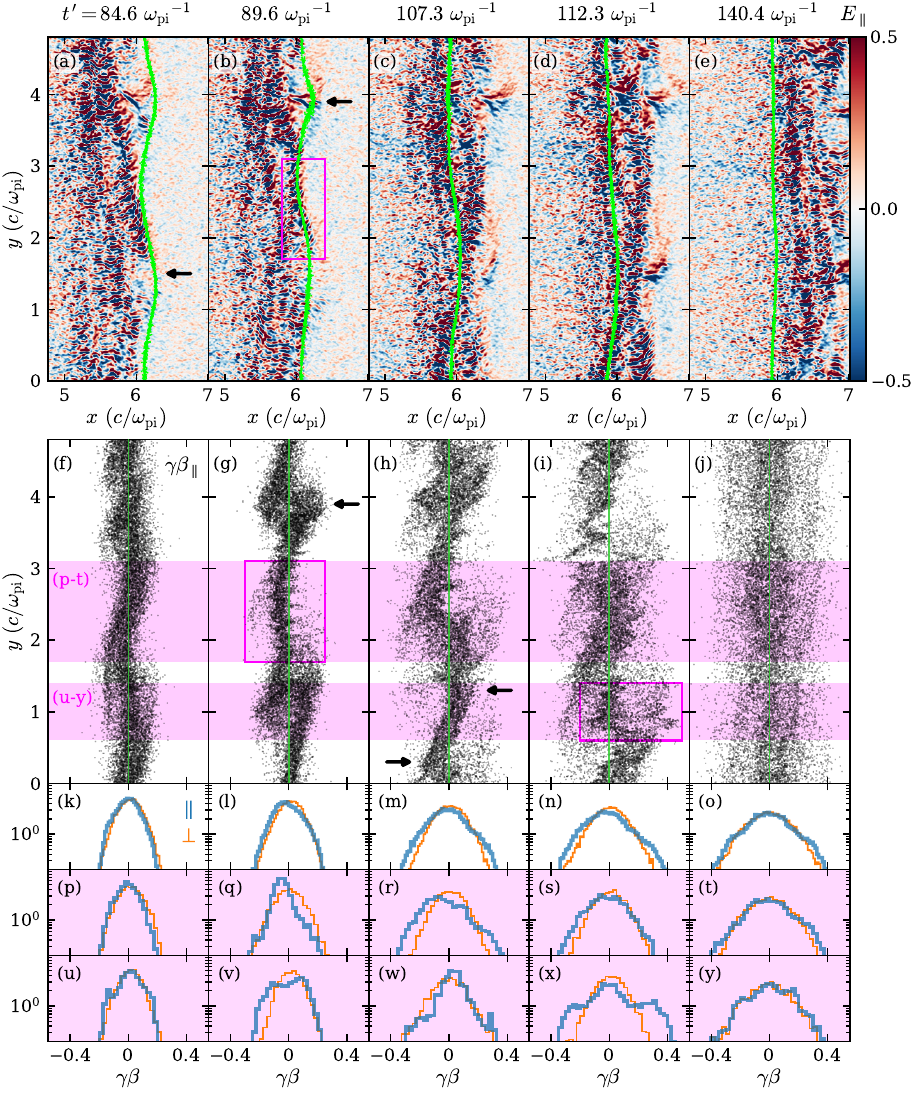}
    \caption{
        Phase space time evolution (left to right) of electron sample in
        $\Mms=3.1$, $\bp=0.25$ shock.
        (a-e):
        $E_\parallel$ normalized to upstream motional field $u_0 B_0/c$
        with electron sample overlaid (green dots).  Colormap saturates on
        small-scale waves.
        (f-j):
        $\gamma \beta_\parallel$--$y$ phase space of electron sample.
        Green vertical lines mark $\gamma \beta_\parallel = 0$.
        (k-o): 1-D $\gamma \beta_{\{\parallel,\perp\}}$ distribution of full electron sample.
        Thick blue curve is $\gamma \beta_\parallel$; orange curve is
        $\gamma \beta_\perp$.
        (p-t): like (k-o), but only electrons within $y=1.7$--$3.1\;c/\ompi$.
        (u-y): like (k-o), but only electrons within $y=0.6$--$1.4\;c/\ompi$.
        Here $\gamma = 1/\sqrt{1-\beta^2}$
        and $\beta_{\{\parallel,\perp\}} = v_{\{\parallel,\perp\}}/c$.
        Arrows and boxes discussed in text.
    }
    \label{fig:distrevol}
\end{figure}

\begin{figure}
    \includegraphics[width=3.375in]{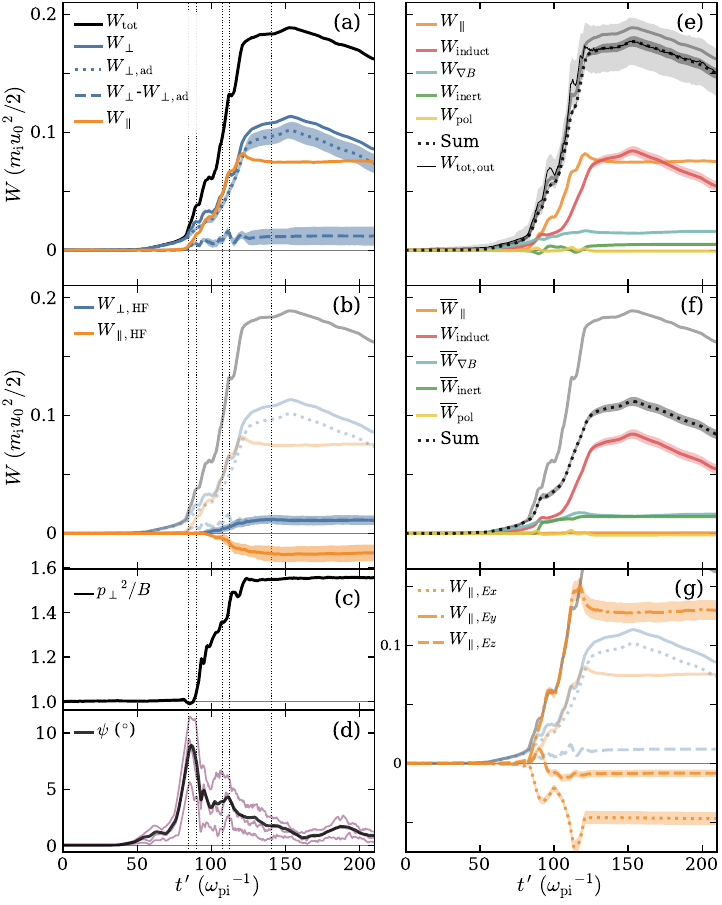}
    \caption{
        Mean work done on electron sample over time, normalized to
        upstream ion drift kinetic energy.  Vertical dotted lines are snapshot
        times in Fig.~\ref{fig:distrevol}.  Faded curves in (b,e,f,g) same as
        (a).  Shaded regions are estimated error on coarse-timestep integrated
        quantities.
        (a): Decomposition into $W_\parallel$, $W_\perp$ and adiabatic work
        $W_{\perp,\mt{ad}}$.
        (b): High-frequency electric field work.
        (c): Particle-averaged adiabatic moment ${p_{\perp}}^2/B$, scaled to
        mean upstream value.
        (d):
        Magnetic field tilt $\psi = \tan^{-1}\left( B_x / \sqrt{{B_y}^2 + {B_z}^2} \right)$
        with respect to $y$-$z$ plane.
        Black curve is particle average $\langle \psi \rangle$;
        purple curves are 25th, 50th (median), and 75th percentiles.
        (e): Parallel, induction, $\nabla B$, inertial, and polarization
        work.  Dark-gray region is error for sum of drift work;
        light-gray region is error for $W_\mt{tot,out}$.
        (f): DC-like drift work, defined in text.
        (g): Parallel work contributions from $E_x$, $E_y$, and $E_z$, as
        defined in text.
        The $y$-axis is offset from (a-b,e-f).
    }
    \label{fig:edotv}
\end{figure}

For further study, we choose the weakest 2-D $\mime=625$, $\bp=0.25$ shock with
significant super-adiabatic heating: our representative $\Mms=3.1$ simulation
(Fig.~\ref{fig:structure}, \ref{fig:machscale}(a)).
We redo this simulation with higher resolution: $\Delta x = \Delta y =
0.05\;\comp$ (keeping $c = 0.45 \Delta x / \Delta t$), $64$ particles per cell
per species, and $64$ current filter passes per timestep.
The current filter approximates a Gaussian with standard deviation
$\abt 5.7$ cells or $0.28 c/\omp$; the filter's half-power cut-off is at wavenumber
$k \approx 3.5 \left(c/\omp\right)^{-1}$, which means 50\% damping at
wavelength $\lambda \approx 1.8 c/\omp$.
We then select all $15898$ electron particles between $x=8.00$--$8.02
\;c/\ompi$ at $t'\equiv t-324\;{\ompi}^{-1} = 0$,
located $4\;c/\ompi=2\;\rLi$ ahead of the shock ramp, and monitor their phase
space evolution (Fig.~\ref{fig:distrevol}) and energy gain
(Fig.~\ref{fig:edotv}) through the shock.
The perpendicular upstream $\vec{B}$ confines particles within a narrow
magnetic flux tube and prevents particle drift from downstream to upstream.

Elongated, ion-scale $E_\parallel$ waves accelerate electrons along $\vec{B}$
in the shock foot and ramp.
These waves have $|E_\parallel|/(u_0 B_0/c) \sim 0.2$--$0.6$ and wavelength
$\lambda_y \sim 2 c/\ompi \sim \rLi$ (Fig.~\ref{fig:distrevol}(a), arrow);
we attribute this $E_\parallel$ to very oblique whistler waves (i.e.,
magnetosonic / lower hybrid branch) with fluctuating $E_x \gg E_y,E_z$ and
$B_y,B_z > B_x$, as identified by prior PIC studies
\citep{matsukiyo2003, hellinger2007, matsukiyo2006, umeda2012-mtsi}.
A stronger bipolar ion-scale $|E_\parallel|/(u_0 B_0/c) \gtrsim 0.5$
(Fig.~\ref{fig:distrevol}(b), arrow) straddles clumps of shock-reflected ions
and also accelerates electrons.
Accelerated electrons appear as coherent deflections in
$\gamma\beta_\parallel$--$y$ phase space (Fig.~\ref{fig:distrevol}(g-h), arrows)
that disrupt and relax via two-stream-like instability.
Local $y$-regions develop asymmetric and transiently unstable $\gamma
\beta_\parallel$ distributions (Fig.~\ref{fig:distrevol}(q,v,x)).
Electron relaxation generates strong and rapid electron-scale $E_\parallel$
waves and phase-space holes with
$\lambda_y \sim c/\omp$
(Fig.~\ref{fig:distrevol}(b,g,i), boxes) \citep[cf.][]{an2019}
Landau damping is evidenced by flattened distributions at
$\gamma\beta_\parallel \sim 0.2$ (Fig.~\ref{fig:distrevol}(k-l)).
Electrons relax to near isotropy by
$t' \sim 140\;{\ompi}^{-1}$
(Fig.~\ref{fig:distrevol}(j,o,t,y)).
Prior 2-D PIC simulations have shown similar two-step $\vec{B}$-parallel
heating in a shock foot setup (periodic interpenetrating beams)
\citep{matsukiyo2006} and in full shocks \citep{umeda2011, umeda2012-mtsi}.

$E_\parallel$ is the main source of non-adiabatic electron heating
(Fig.~\ref{fig:edotv}(a)).
We decompose the sample electrons' mean energy
gain $W_\mt{tot}$ into parallel and perpendicular work,
$W_\parallel = -e \langle \int E_\parallel v_\parallel \dtl t \rangle$ and
$W_\perp = -e \langle \int E_\perp v_\perp \dtl t \rangle$,
integrated for every particle over
every code timestep $\Delta t$ such that $W_\mt{tot} = W_\parallel + W_\perp$.
Angle brackets are particle averages.
We estimate the adiabatic heating as
$W_{\perp,\mt{ad}} = \langle \sum_n (\gamma_{n \to n+1,\mt{ad}} - \gamma_n) \me c^2 \rangle$,
where
\begin{equation}
    \gamma_{n \to n+1,\mt{ad}}
    = \sqrt{1 + \left(\gamma\beta_\parallel\right)^2_n
        + \left(\gamma\beta_\perp\right)^2_n \left(B_{n+1}/B_n\right) }
    \label{eq:adgamma}
\end{equation}
captures electron heating from compression between timesteps $n$ and $n+1$.

In Eqn.~(\ref{eq:adgamma}),
we assume $(\gamma \beta_\perp)^2_n/B_n$ and $(\gamma\beta_\parallel)_n$
are constant during compression;
$\gamma$, $\beta_\parallel$, and $\beta_\perp$ are evaluated in
the electron fluid's rest frame.
The sum in $W_{\perp,\mt{ad}}$ uses a coarse output timestep
$\Delta t_\mt{out} = 400 \; \Delta t = 9\;{\omp}^{-1}$.

The non-adiabatic perpendicular work can be explained by high-frequency
scattering of electron parallel energy into perpendicular energy
(Fig.~\ref{fig:edotv}(b)).
We Fourier-space filter $E_x$, $E_y$, and $E_z$ to isolate wavenumbers
$k_y > \left(c/\omp\right)^{-1}$ and then construct $E_{\perp,\mt{HF}}$ and
$E_{\parallel,\mt{HF}}$
by projecting the Fourier-filtered fields onto local $\vec{B}$.
Then,
$W_{\perp,\mt{HF}} = - e \langle \sum E_{\perp,\mt{HF}} v_{\perp} \Delta t_\mt{out} \rangle$
and $W_{\parallel,\mt{HF}} = - e \langle \sum E_{\parallel,\mt{HF}} v_{\parallel} \Delta t_\mt{out} \rangle$.
We find that $W_{\perp,\mt{HF}}$ and $W_{\perp} - W_{\perp,\mt{ad}}$ agree to
$\sim 10\%$, suggesting that non-adiabatic $W_\perp$ comes from electron-scale
scattering of parallel energy.
Exact agreement is not expected due to the coarse timestep $\Delta t_\mt{out}$
and the arbitrary $k_y$ cut.

The particle-averaged adiabatic moment ${p_\perp}^2/B$ grows in steps that
correlate with increases in $W_\mt{tot}$ and $W_\parallel$.
Bulk acceleration at $t'=84.6\;{\ompi}^{-1}$ and $t'=107.3\;{\ompi}^{-1}$
coincides with momentarily constant ${p_\perp}^2/B$ and increasing
$W_\parallel$ prior to a scattering episode.
Then, ${p_\perp}^2/B$ increases during strong electron scattering at
$t'=89.6\;{\ompi}^{-1}$ and $112.3\;{\ompi}^{-1}$ while $W_\parallel$ flattens
off (Fig.~\ref{fig:distrevol}, Fig.~\ref{fig:edotv}(a,c)).

Fig.~\ref{fig:edotv}(e) shows mean work from grad B, inertial, and polarization
drifts, as well as $\partial B/\partial t$ induction work; see
\citet{northrop1961,northrop1963,goodrich1984,dahlin2014,rowan2019}.
Each $W_\mt{drift} = - e \langle \sum \vec{E} \cdot \vec{v}_\mt{drift} \Delta t_\mt{out} \rangle$,
where $\vec{v}_\mt{drift}$ is one of:
\[
    \vec{v}_{\{\nabla B, \mt{inert}, \mt{pol}\}}
        = - \frac{\gamma \me c}{e B} \hat{\vec{b}} \times \left\{
            \frac{{v_\perp}^2}{2 B} \nabla B
            \boldsymbol{,}\, v_\parallel \frac{\mt{d}\hat{\vec{b}}}{\mt{d}t}
            \boldsymbol{,}\, \frac{\mt{d}\vec{v}_E}{\mt{d}t}
        \right\} ,
\]
with $\gamma$ and $v_\perp$ evaluated in the electron fluid's rest frame.
We take $\vec{v}_E = \langle c\vec{E}\times\vec{B}/B^2 \rangle$ to reduce
noise; otherwise, the $\vec{v}_\mt{drift}$ terms use $\vec{E}$ and $\vec{B}$
fields seen by individual particles.
The $\mt{d}/\mt{d}t$ terms are one-sided finite differences; e.g.,
$\mt{d}\vec{v}_E/\mt{d}t = [(\vec{v}_E)_{n+1} - (\vec{v}_E)_{n}]/\Delta t_\mt{out}$.
And, $W_\mt{induct} = \gamma\me{v_\perp}^2 (\partial B/\partial t)/(2 B)$,
with $\partial B/\partial t = [B_{n+1}(\vec{r}_n) - B_{n-1}(\vec{r}_n)]/(2\Delta t_\mt{out})$
and $\vec{r}_n$ the particle position at timestep $n$.
We find that grad B drift and induction together give fluid-like adiabatic compression.
Inertial and polarization drifts give less work, but some other electron samples
have $W_\mt{inert}$ comparable to $W_{\nabla B}$ (Appendix~\ref{app:more}).
We compare the summed drifts to
$W_\mt{tot,out} = -e \langle \sum \vec{E}\cdot\vec{v}\Delta t_\mt{out}\rangle$.
We conclude that $W_\mt{tot}$ agrees with both the summed drift work and
$W_\mt{tot,out}$, given uncertainty from both the guiding-center drift
approximation and the coarse integration timestep.

Earlier, we argued that DC heating alone may not explain all super-adiabatic
heating in our fiducial 2-D shock, based on downstream volume-averaged $\TeTi$
(Figs.~\ref{fig:structure}(c), \ref{fig:machscale}(b)).
So, Fig.~\ref{fig:edotv}(f) estimates DC-like work as
$\overline{W}_\mt{drift} = - e \sum \langle\vec{E}\rangle \cdot \langle\vec{v}_\mt{drift}\rangle \Delta t_\mt{out}$
and
$\overline{W}_\parallel = - e \sum \langle E_\parallel\rangle \langle v_\parallel \rangle \Delta t_\mt{out}$.
The $\vec{E}$ average removes waves along $\hat{\vec{y}}$ to keep only 1-D-like
shock fields.
The $\vec{v}_\mt{drift}$ average gives a mean drift trajectory and mostly discards gyration.
The DC-like parallel work $\overline{W}_\parallel$ goes to zero,
and the DC-like contribution to super-adiabatic heating appears small.
Fluid-like adiabatic compression is preserved in $W_\mt{induct} + \overline{W}_{\nabla B}$.
Fig.~\ref{fig:edotv}(g) separates $E_x$, $E_y$, and $E_z$ contributions to
$W_\parallel$ as
$W_{\parallel,Ei} = - e \langle \sum E_i b_i v_\parallel \Delta t_\mt{out} \rangle$,
where $i=x,y,z$ and $b_i$ is the $i$-th component of $\hat{\vec{b}}$.
$E_y$ gives parallel heating, whereas $E_x$ and $E_z$ cause parallel cooling.

The quantities $W_{\perp,\mt{ad}}$, $W_{\{\parallel,\perp\},\mt{HF}}$,
$W_\mt{drift}$, $W_\mt{tot,out}$, $\overline{W}_\mt{drift}$,
$\overline{W}_\parallel$, and $W_{\parallel,Ei}$
are integrated with coarse timestep $\Delta t_\mt{out}$ and
converged at the $\abt 10\%$ level.
The error regions in Fig.~\ref{fig:edotv}(a,b,e) are defined in
Appendix~\ref{app:edotvconv}.

\section{Conclusion}

We have measured $\TeTi$ in 2-D PIC simulations of perpendicular shocks to
inform models of astrophysical systems lacking direct $\Te$ or $\Ti$
measurements.
In a $\Mms=3.1$, $\bp=0.25$ rippled shock, quasi-static DC fields
provide fluid-like adiabatic heating, and most super-adiabatic heating is
from ion-scale $E_\parallel$ waves.

\acknowledgments

Xinyi Guo shared and provided helpful assistance for some software used to
perform and analyze these simulations.
Alex Bergier and colleagues provided excellent assistance with Columbia's
Habanero cluster.
Adam Masters and Parviz Ghavamian kindly shared Saturn bow shock data.
We thank Matthew W.~Abruzzo, Luca Comisso, Greg Howes, Anatoly Spitkovsky,
Vassilis Tsiolis, and Lynn B.~Wilson III for discussion.
We thank the anonymous referees for integral comments and critiques.
LS and AT were supported by the Sloan Fellowship to LS, NASA ATP-80NSSC20K0565, and NSF AST-1716567.
Some work was done at UCSB KITP, which is supported by NSF PHY-1748958.
Simulations were run on Habanero (Columbia University), Edison (NERSC), and
Pleiades (NASA HEC).
Columbia University's Shared Research Computing Facility is supported by NIH
Research Facility Improvement Grant 1G20RR030893-01 and the New York State
Empire State Development, Division of Science Technology and Innovation
(NYSTAR) Contract C090171.
NERSC is a U.S.~Department of Energy Office of Science User Facility operated
under Contract DE-AC02-05CH11231.
The NASA HEC Program is part of the NASA Advanced Supercomputing (NAS) Division
at Ames Research Center.

\facility{NERSC, Pleiades}

\listofchanges

\bibliographystyle{aasjournal}
\bibliography{lechtg}

\appendix

\section{More Views of Electron Sample Heating}\label{app:more}

\begin{figure}
    \begin{interactive}{animation}{{mov_distr-on-fld}.mp4}
    \includegraphics[width=\columnwidth]{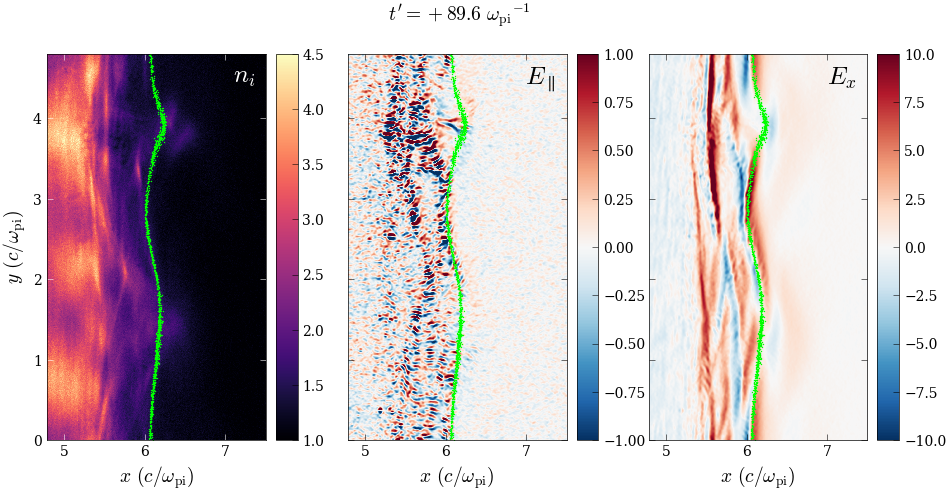}
    \end{interactive}
    \caption{
        Movie still: electron sample at $t'=89.6\;{\ompi}^{-1}$ plotted
        over ion density $n_\mathrm{i}$, parallel electric field $E_\parallel$,
        and electric field component $E_x$; same sample from
        Figs.~\ref{fig:distrevol} and \ref{fig:edotv}.
        An animation of this figure from $t' = 54$ to $180\;{\ompi}^{-1}$.
        is in the online journal.
        The ion density $n_\mathrm{i}$ is scaled to upstream density $n_0$, and
        the electric field components are scaled to upstream motional electric
        field magnitude $u_0 B_0 / c$.
        The $E_\parallel$ colormap spans $[-1,+1]$ and saturates on small-scale
        waves, despite being a wider range than Fig.~\ref{fig:distrevol}.
    }
    \label{fig:mov-distr-on-fld}
\end{figure}

\begin{figure}
    \begin{interactive}{animation}{{mov_distrevol}.mp4}
    \includegraphics[width=\columnwidth]{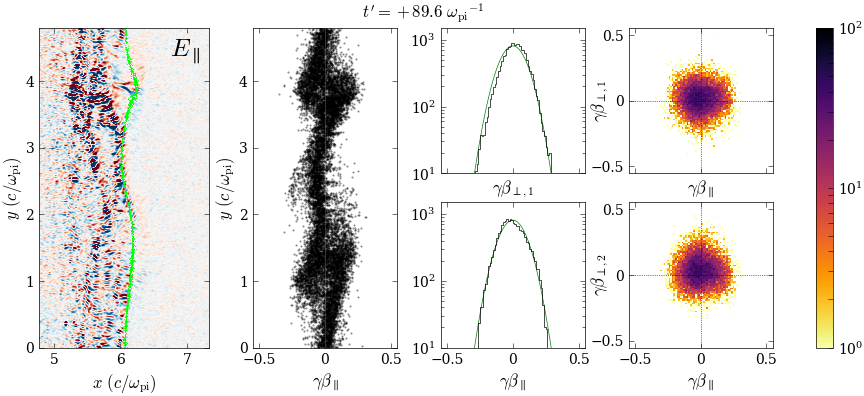}
    \end{interactive}
    \caption{
        Movie still: electron sample phase space at $t'=89.6\;{\ompi}^{-1}$,
        with more detail than Fig.~\ref{fig:distrevol}.
        An animation of this figure from $t' = 54$ to $180\;{\ompi}^{-1}$.
        is in the online journal.
        The momentum component $\gamma \beta_{\perp,1}$ is the projection along
        $(\hat{\vec{b}} \times -\hat{\vec{x}}) \times \hat{\vec{b}}$,
        and the component $\gamma \beta_{\perp,2}$ is the projection along
        $\hat{\vec{b}} \times -\hat{\vec{x}}$.
        Because the local magnetic field unit vector $\hat{\vec{b}}$ mostly orients along
        $\hat{\vec{y}}$, the components $\perp,1$ and $\perp,2$ roughly correspond to
        $-\hat{\vec{x}}$ and $+\hat{\vec{z}}$ so that ($\perp,1$; $\perp,2$; $\parallel$) form a
        right-handed coordinate system.
        In the 1-D phase space plots, the light green curve is an isotropic
        Maxwell-J\"{u}ttner distribution with same mean energy as the
        electrons.
        The $E_\parallel$ colormap is the same as in Fig.~\ref{fig:mov-distr-on-fld}.
    }
    \label{fig:mov-distrevol}
\end{figure}

Figs.~\ref{fig:mov-distr-on-fld} and \ref{fig:mov-distrevol} present two
movies, available online, of our electron sample traversing the shock front.

\begin{figure}
    \includegraphics[width=\columnwidth]{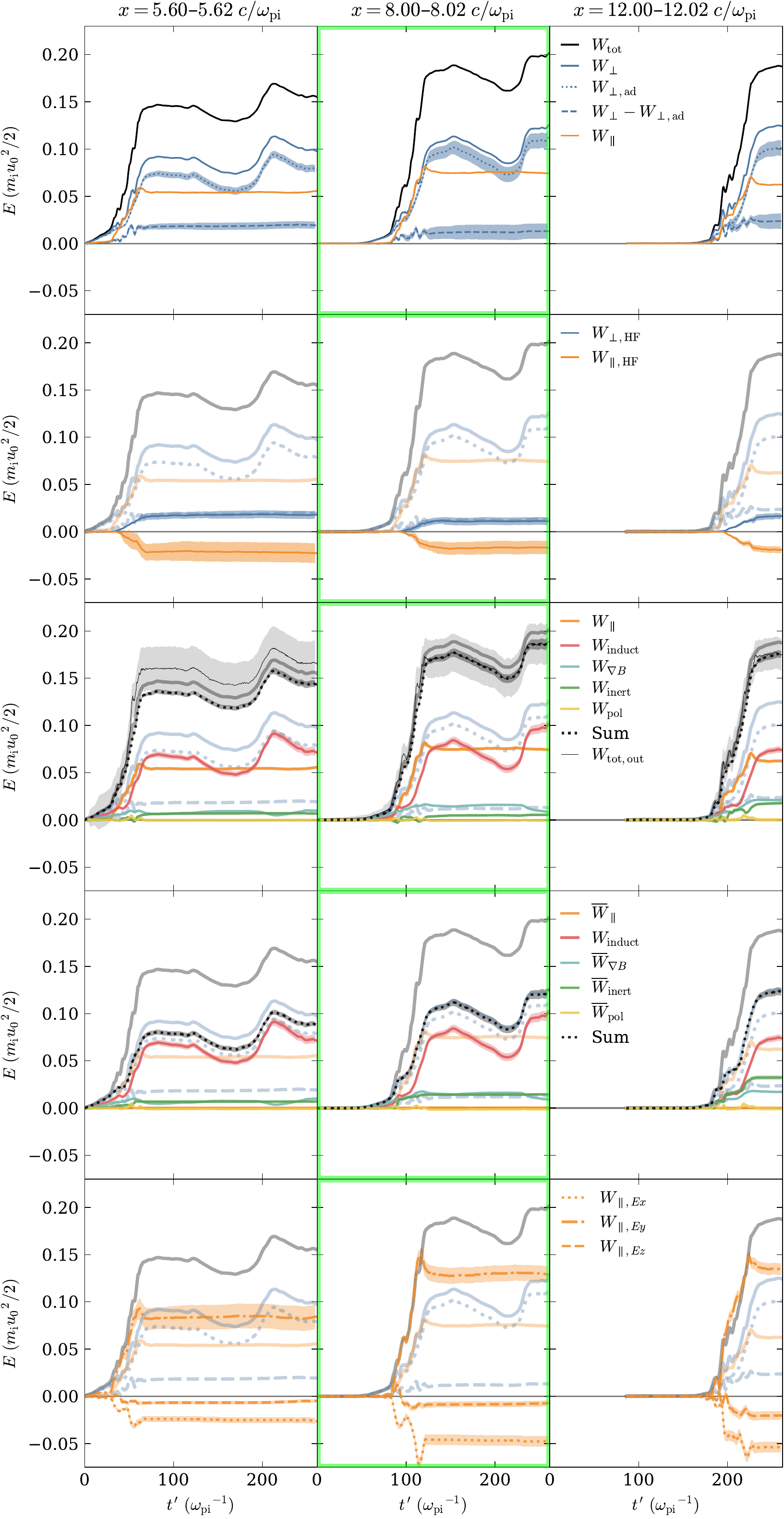}
    \caption{
        Like Fig.~\ref{fig:edotv}(a-b,e-g), but for multiple electron
        samples traversing the shock at different times.
        Left and right: first and last samples tracked.
        Middle (green box): sample shown in Figs.~\ref{fig:distrevol} and \ref{fig:edotv}.
        Column titles indicate the sample's location at $t' = 0 \; {\ompi}^{-1}$.
        The complete figure (17 samples) is available in the online journal.
    }
    \label{fig:edotvmore}
\end{figure}

In Fig.~\ref{fig:edotvmore}, we show the work decomposition from
Fig.~\ref{fig:edotv} for many electron samples.
At $t' = 0 \; {\ompi}^{-1}$, we selected all electrons in the regions
$x\in[5.60,5.62]c/\ompi$, $x\in[6.00,6.02]c/\ompi$, and so on with even spacing
$0.4 c/\ompi$ to get seventeen electron particle samples of similar size.

\figsetstart
\figsetnum{\ref{fig:edotvmore}}
\figsettitle{Electron Work Decomposition}
\figsetend

\section{Convergence in Electron Work Summation}\label{app:edotvconv}

Several quantities in Figure~\ref{fig:edotv} are summed with a coarse timestep
$\Delta t_\mt{out} = 400 \Delta t = 9\;{\omp}^{-1}$, namely:
$W_\mathrm{\perp,ad}$,
$W_\mathrm{\perp} - W_\mathrm{\perp,ad}$,
$W_\mathrm{\perp,HF}$,
$W_\mathrm{\parallel,HF}$,
$W_\mt{induct}$,
$W_\mt{\nabla B}$,
$W_\mt{inert}$,
$W_\mt{pol}$,
$W_\mt{tot,out}$,
$\overline{W}_\mt{\nabla B}$,
$\overline{W}_\mt{inert}$,
$\overline{W}_\mt{pol}$,
$\overline{W}_\parallel$,
$W_{\parallel,Ex}$,
$W_{\parallel,Ey}$,
and $W_{\parallel,Ez}$.
To check convergence, we downsample in time each quantity's summand by $4$ and compute
an error $\delta f$ at discrete time $t_n$ as:
\begin{equation}
    \delta f(t_n) = \max \left\{ \left| f(t_m) - f_{1/4}(t_m) \right| \mid 0 \le m \le n \right\}
    \label{eq:error}
\end{equation}
where $f_{1/4}$ is the $4\times$ downsampled version of $f$.
Thus $\delta f$ is strictly non-decreasing with $t_n$.
The error regions defined by Eq.~\ref{eq:error} are plotted as shaded areas in
Figs.~\ref{fig:edotv} and \ref{fig:edotvmore}.

\begin{figure}
    \includegraphics[width=\columnwidth]{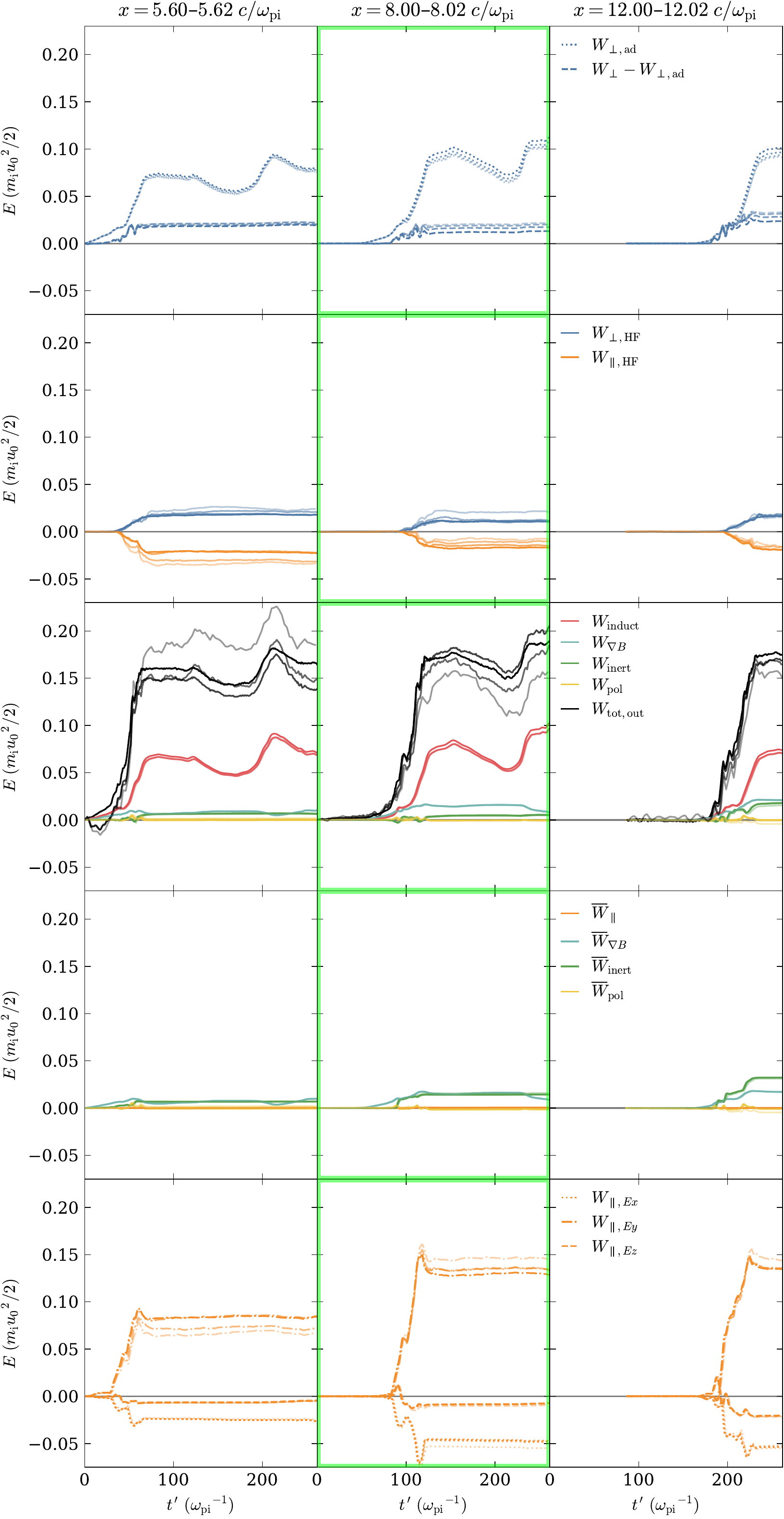}
    \caption{
        Like Fig.~\ref{fig:edotvmore}, but showing
        the numerical convergence of all quantities integrated with the coarse
        timestep $\Delta t_\mt{out}$, listed in text.
        For each quantity, we increase the sample spacing $\Delta t_\mathrm{out}$ by
        $2\times$, $4\times$, and $8\times$ and plot the down-sampled integration with
        progressively decreasing opacity.
        The complete figure (17 samples) is available in the online journal.
    }
    \label{fig:edotvconvmore}
\end{figure}

Fig.~\ref{fig:edotvconvmore} shows curves with $2,4,8\times$ downsampling for
the seventeen distinct electron samples of Fig.~\ref{fig:edotvmore}, including
the sample shown in Fig.~\ref{fig:edotv}.

\figsetstart
\figsetnum{\ref{fig:edotvconvmore}}
\figsettitle{Electron Work Convergence}
\figsetend

\section{Transverse Width Convergence}\label{app:widthconv}

\begin{figure*}
    \plotone{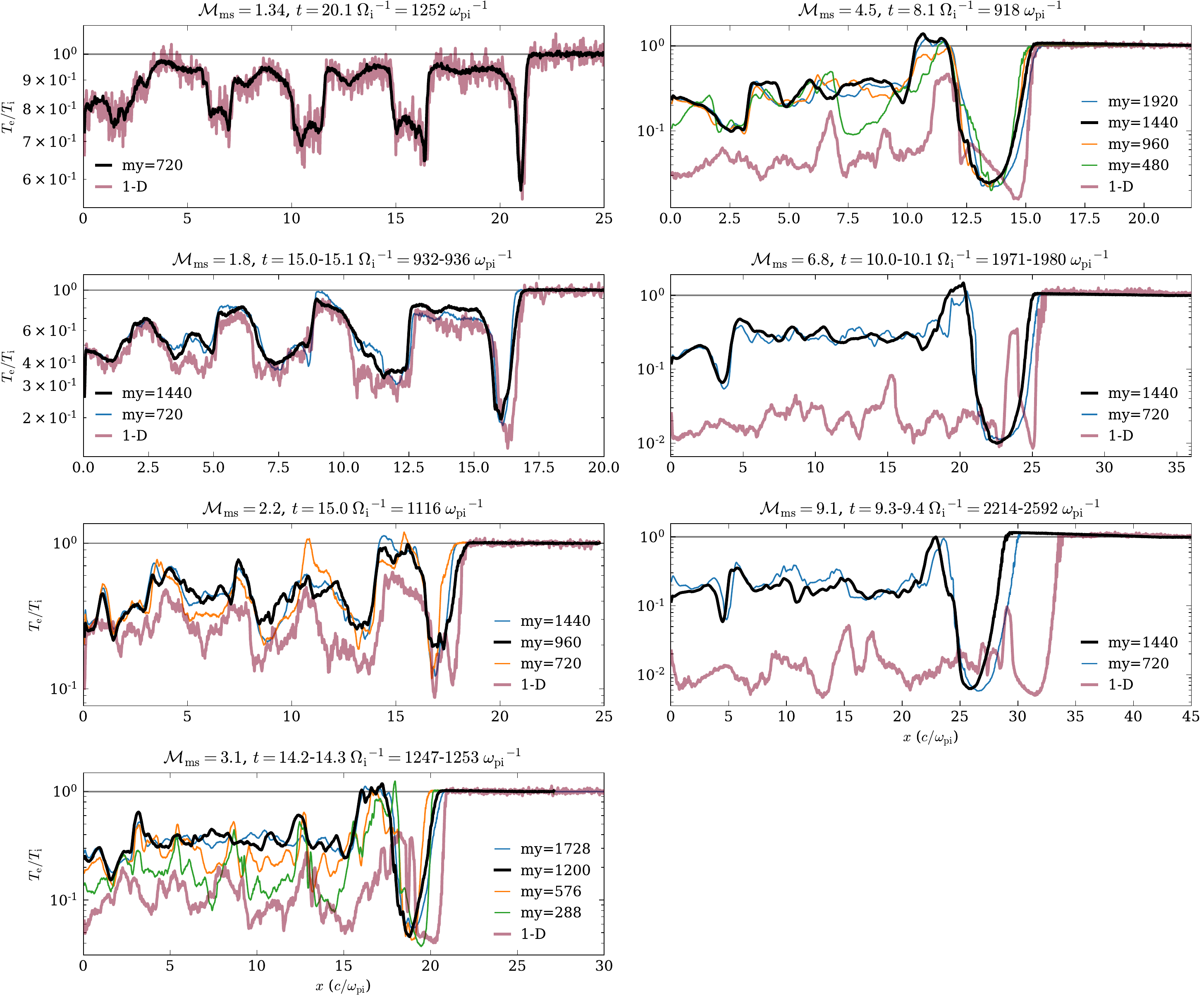}
    \caption{
    Fiducial 2-D simulations are converged with respect to transverse width.
    Black curves are fiducial 2-D $\mime=625$, $\bp=0.25$ simulations from
    Fig.~\ref{fig:machscale}(a).
    Colored curves are same shock parameters with varying input \texttt{my}, as
    defined in Table~\ref{tab:param}.
    A time range is given because simulation output times do not match exactly.
    }
    \label{fig:width}
\end{figure*}

Fig.~\ref{fig:width} shows that our fiducial 2-D simulations are converged with
respect to transverse width.
Most of the varying transverse width simulations are \emph{not} listed in
Table~\ref{tab:param}.
For $\Mms=9.1$ only, the 1-D simulation uses a slightly higher upstream
temperature than the 2-D simulations, so the ratio $\Omci/\ompi$ differs
between 1-D and 2-D.  In this case, we matched times based on ${\Omci}^{-1}$
rather than ${\ompi}^{-1}$.

\section{Simulation Parameters}\label{app:param}

Table~\ref{tab:param} contains input parameters, derived shock parameters, run
durations, and downstream temperature measurements for all simulations in our
manuscript.
The first row is the high-resolution run used in Figs.~\ref{fig:distrevol} and
\ref{fig:edotv}; the remaining rows are presented in Fig.~\ref{fig:machscale}.
A machine-readable version of Table~\ref{tab:param}, in comma-separated value
(CSV) ASCII, is available in the online journal.
Below, we define all table columns.

\begin{itemize}

\item
\texttt{mi\_me} is the ion-electron mass ratio $\mime$.

\item
\texttt{theta} and \texttt{phi} specify the upstream magnetic field
orientation, measured in the simulation frame.
$\theta$ is the angle between $\vec{B}$ and the $x$-coordinate axis.
$\varphi$ is the angle between the $y$-$z$ plane projection of $\vec{B}$ and
the $z$-coordinate axis.
To visualize these angles, see Fig.~1 of \citet{guo2014-accel}, but note that
their $\varphi_\mt{B} = \pi/2 - \varphi$ is the complement of our $\varphi$.
For all our simulations, $\theta$ corresponds to the angle between $\vec{B}$
and shock normal.
The 2-D simulations with in-plane $\vec{B}$
have $\theta=90^{\circ}$ and $\varphi=90^{\circ}$.
The 2-D simulations with out-of-plane $\vec{B}$ (i.e., $\vec{B}$ along $\hat{\vec{z}}$)
have $\theta=90^{\circ}$ and $\varphi=0^{\circ}$.
The 1-D simulations with oblique $\vec{B}$ have
have $\theta<90^{\circ}$.

\item
\texttt{my} and \texttt{mz} are the numbers of grid cells along $\hat{\vec{y}}$ and
$\hat{\vec{z}}$.  Our 2-D simulations have $\mathtt{mz}=1$, and 1-D simulations have
$\mathtt{my}=\mathtt{mz}=1$.

\item
\texttt{betap}, \texttt{Ms}, \texttt{Ma}, and \texttt{Mms} are the shock plasma
beta $\bp$, sonic Mach number $\Ms$, Alfv\'{e}n Mach number $\Ma$, and fast
magnetosonic Mach number $\Mms$.
These numbers are \emph{derived} from TRISTAN-MP input parameters
\texttt{sigma}, \texttt{delgam}, and \texttt{u0} (defined just below).
First, the total plasma beta is:
\[
    \bp = \frac{
        4 \gamma_0 \Delta\gamma_\mathrm{i}
    }{
        \sigma \left(\gamma_0 - 1\right) \left(1 + \me/\mi\right)
    }
\]
where $\gamma_0 = 1/\sqrt{1 - \left(u_0/c\right)^2}$ is the Lorentz factor of
the upstream flow in the simulation frame.
The sonic Mach number depends on the upstream plasma speed in the shock's rest
frame:
\[
    \Ms = \frac{u_\mathrm{sh}}{c_\mathrm{s}} = \frac{u_\mathrm{0}}{\left(1 - 1/r\left(\Ms\right)\right) c_\mathrm{s}}
\]
and we solve this implicit expression for $\Ms$ (and thus also $u_\mathrm{sh}$)
using an input $u_\mathrm{0}$ and assumed fluid adiabatic index $\Gamma = 2$
(note that $\Gamma$ enters into both the Rankine-Hugoniot expression for MHD
shock compression ratio $r$ and the sound speed $c_\mathrm{s}$).
Once $\Ms$ and $u_\mathrm{sh}$ are known, $\Ma$ and $\Mms$ are known as well.
This procedure for estimating shock parameters is taken directly from
\citet{guo2017}.

\item
\texttt{sigma} is the magnetization, a ratio of upstream magnetic and kinetic
enthalpy densities:
\[
\texttt{sigma} = \sigma \equiv \frac{{B_0}^2}{4\pi \left(\gamma_0 - 1\right) \left(\mi+\me\right) n_0 c^2} .
\]

\item
\texttt{delgam} is the upstream plasma temperature, scaled by ion rest energy:
\[
\texttt{delgam} = \Delta \gamma_\mathrm{i} \equiv \frac{\kB T_0}{\mi c^2} .
\]

\item
\texttt{u0} is the upstream plasma velocity, scaled by speed of light:
\[
\texttt{u0} \equiv u_0/c
\]

\item
\texttt{ppc0} is number of particles (both electrons and ions) per cell in the
upstream plasma.

\item
\texttt{c\_omp} is the number of grid cells per electron skin depth $c/\omp$.

\item
\texttt{ntimes} is the number of current filter passes.

\item
\texttt{dur} is the simulation duration in units of upstream ion cyclotron time
${\Omci}^{-1}$.

\item
\texttt{Te\_Ti} is our measurement of downstream temperature ratio $\TeTi$.
As described in the main text, we manually choose a downstream region that is
minimally affected by the left-side reflecting wall and the right-side shock
front relaxation.
Our measurement of $\TeTi$ uses downsampled grid output of the particle
temperature tensor; however, the temperature tensor itself is calculated for
each grid cell using the full particle distribution in a $5^N$ cell region,
where $N \in \{1,2,3\}$ is the domain dimensionality.

\item
\texttt{Te\_Ti\_std} is the standard deviation of $\TeTi$ within the downstream
region that we consider.
Like \texttt{Te\_Ti}, downsampled grid output is used for this estimate.

\item
\texttt{Te} and \texttt{Ti} are the downstream electron and ion temperatures
scaled to their respective rest masses; i.e., $\kB \Te/(\me c^2)$ and
$\kB \Ti/(\mi c^2)$.
We measure all of \texttt{Te}, \texttt{Ti}, and \texttt{Te\_Ti} in the same
manually-chosen downstream region.

\end{itemize}

\begin{longrotatetable}
\begin{deluxetable*}{rrrrrcrrrcccrrrrcccc}
\tabletypesize{\scriptsize}
\movetabledown=0.2in
\tablecaption{
    Simulation input parameters, derived shock parameters, run duration, and
    downstream temperature measurements.
    Columns are defined in Appendix~\ref{app:param}.
    \label{tab:param}
}
\tablehead{
    \multicolumn{1}{l}{\texttt{mi\_me}}  
      & \multicolumn{1}{r}{\texttt{theta}}
      & \multicolumn{1}{r}{\texttt{phi}}
      & \multicolumn{1}{r}{\texttt{my}}
      & \multicolumn{1}{r}{\texttt{mz}}
      & \multicolumn{1}{c}{\texttt{betap}}
      & \multicolumn{1}{r}{\texttt{Ms}}
      & \multicolumn{1}{r}{\texttt{Ma}}
      & \multicolumn{1}{r}{\texttt{Mms}}
      & \multicolumn{1}{c}{\texttt{sigma}}
      & \multicolumn{1}{c}{\texttt{delgam}}
      & \multicolumn{1}{c}{\texttt{u0}}
      & \multicolumn{1}{r}{\texttt{ppc0}}
      & \multicolumn{1}{r}{\texttt{c\_omp}}
      & \multicolumn{1}{r}{\texttt{ntimes}}
      & \multicolumn{1}{r}{\texttt{dur}}
      & \multicolumn{1}{c}{\texttt{Te\_Ti}}
      & \multicolumn{1}{c}{\texttt{Te\_Ti\_std}}
      & \multicolumn{1}{c}{\texttt{Te}}
      & \multicolumn{1}{c}{\texttt{Ti}} \\
}
\startdata
625 & 90 & 90 & 2400 & 1 & 0.250 & 6.86 & 3.43 & 3.07 & 4.7854E$-$1 & 8.0944E$-$6 & 2.3245E$-$2 & 128 & 20 & 64 & 6.7 & \nodata & \nodata & \nodata & \nodata \\
\hline
625 & 90 & 90 & 720 & 1 & 0.251 & 3.00 & 1.50 & 1.34 & 1.0117E$+$1 & 1.6189E$-$5 & 7.1502E$-$3 & 32 & 10 & 32 & 20.1 & 8.53E$-$1 & 1.38E$-$1 & 1.29E$-$2 & 2.43E$-$5 \\
625 & 90 & 90 & 1440 & 1 & 0.250 & 4.00 & 2.00 & 1.79 & 2.3774E$+$0 & 1.6189E$-$5 & 1.4749E$-$2 & 32 & 10 & 32 & 15.0 & 5.59E$-$1 & 2.33E$-$1 & 1.74E$-$2 & 4.96E$-$5 \\
625 & 90 & 90 & 960 & 1 & 0.250 & 5.00 & 2.50 & 2.24 & 1.1237E$+$0 & 1.1332E$-$5 & 1.7949E$-$2 & 32 & 10 & 32 & 20.1 & 4.52E$-$1 & 1.69E$-$1 & 1.96E$-$2 & 6.92E$-$5 \\
625 & 90 & 90 & 1200 & 1 & 0.250 & 6.86 & 3.43 & 3.07 & 4.7854E$-$1 & 8.0944E$-$6 & 2.3245E$-$2 & 32 & 10 & 32 & 14.2 & 3.65E$-$1 & 1.28E$-$1 & 2.83E$-$2 & 1.24E$-$4 \\
625 & 90 & 90 & 1440 & 1 & 0.250 & 9.99 & 4.99 & 4.47 & 1.9968E$-$1 & 4.8566E$-$6 & 2.7873E$-$2 & 32 & 10 & 32 & 12.1 & 3.18E$-$1 & 1.22E$-$1 & 3.65E$-$2 & 1.83E$-$4 \\
625 & 90 & 90 & 1440 & 1 & 0.250 & 15.19 & 7.60 & 6.79 & 8.1402E$-$2 & 1.6189E$-$6 & 2.5205E$-$2 & 32 & 10 & 32 & 10.0 & 2.97E$-$1 & 1.22E$-$1 & 2.80E$-$2 & 1.51E$-$4 \\
625 & 90 & 90 & 1440 & 1 & 0.250 & 20.37 & 10.18 & 9.11 & 4.4398E$-$2 & 8.0944E$-$7 & 2.4133E$-$2 & 32 & 10 & 32 & 9.8 & 1.98E$-$1 & 5.99E$-$2 & 1.83E$-$2 & 1.47E$-$4 \\
\hline
625 & 90 & 0 & 1440 & 1 & 0.251 & 3.00 & 1.50 & 1.34 & 1.0117E$+$1 & 1.6189E$-$5 & 7.1502E$-$3 & 32 & 10 & 32 & 19.0 & 8.62E$-$1 & 1.20E$-$1 & 1.30E$-$2 & 2.42E$-$5 \\
625 & 90 & 0 & 1440 & 1 & 0.250 & 4.00 & 2.00 & 1.79 & 2.3774E$+$0 & 1.6189E$-$5 & 1.4749E$-$2 & 32 & 10 & 32 & 19.9 & 4.71E$-$1 & 1.78E$-$1 & 1.60E$-$2 & 5.42E$-$5 \\
625 & 90 & 0 & 1440 & 1 & 0.250 & 5.00 & 2.50 & 2.24 & 1.1237E$+$0 & 1.1332E$-$5 & 1.7949E$-$2 & 32 & 10 & 32 & 15.0 & 2.49E$-$1 & 1.28E$-$1 & 1.24E$-$2 & 7.93E$-$5 \\
625 & 90 & 0 & 1440 & 1 & 0.250 & 7.00 & 3.50 & 3.13 & 4.5493E$-$1 & 8.0944E$-$6 & 2.3840E$-$2 & 32 & 10 & 32 & 15.2 & 1.07E$-$1 & 5.45E$-$2 & 1.02E$-$2 & 1.54E$-$4 \\
625 & 90 & 0 & 1440 & 1 & 0.250 & 10.00 & 5.00 & 4.47 & 1.9912E$-$1 & 4.8566E$-$6 & 2.7912E$-$2 & 32 & 10 & 32 & 9.8 & 4.48E$-$2 & 2.30E$-$2 & 6.82E$-$3 & 2.44E$-$4 \\
625 & 90 & 0 & 1440 & 1 & 0.250 & 15.00 & 7.50 & 6.71 & 8.3580E$-$2 & 1.6189E$-$6 & 2.4874E$-$2 & 32 & 10 & 32 & 8.5 & 2.54E$-$2 & 2.43E$-$2 & 3.11E$-$3 & 1.96E$-$4 \\
\hline
625 & 90 & 90 & 1 & 1 & 0.251 & 3.00 & 1.50 & 1.34 & 1.0117E$+$1 & 1.6189E$-$5 & 7.1502E$-$3 & 512 & 10 & 32 & 40.0 & 8.52E$-$1 & 1.05E$-$1 & 1.32E$-$2 & 2.47E$-$5 \\
625 & 90 & 90 & 1 & 1 & 0.250 & 4.00 & 2.00 & 1.79 & 2.3774E$+$0 & 1.6189E$-$5 & 1.4749E$-$2 & 512 & 10 & 32 & 40.0 & 4.28E$-$1 & 1.86E$-$1 & 1.58E$-$2 & 5.92E$-$5 \\
625 & 90 & 90 & 1 & 1 & 0.250 & 5.00 & 2.50 & 2.24 & 1.1237E$+$0 & 1.1332E$-$5 & 1.7949E$-$2 & 512 & 10 & 32 & 25.2 & 2.36E$-$1 & 8.40E$-$2 & 1.24E$-$2 & 8.40E$-$5 \\
625 & 90 & 90 & 1 & 1 & 0.250 & 7.00 & 3.50 & 3.13 & 4.5493E$-$1 & 8.0944E$-$6 & 2.3840E$-$2 & 512 & 10 & 32 & 25.0 & 1.01E$-$1 & 3.87E$-$2 & 1.02E$-$2 & 1.61E$-$4 \\
625 & 90 & 90 & 1 & 1 & 0.250 & 10.00 & 5.00 & 4.47 & 1.9912E$-$1 & 4.8566E$-$6 & 2.7912E$-$2 & 512 & 10 & 32 & 15.2 & 5.02E$-$2 & 3.44E$-$2 & 7.03E$-$3 & 2.24E$-$4 \\
625 & 90 & 90 & 1 & 1 & 0.250 & 15.00 & 7.50 & 6.71 & 8.3580E$-$2 & 1.6189E$-$6 & 2.4874E$-$2 & 1024 & 10 & 32 & 15.1 & 2.38E$-$2 & 1.57E$-$2 & 2.85E$-$3 & 1.92E$-$4 \\
625 & 90 & 90 & 1 & 1 & 0.250 & 20.00 & 10.00 & 8.94 & 4.6080E$-$2 & 1.1332E$-$6 & 2.8027E$-$2 & 2048 & 10 & 32 & 10.1 & 1.35E$-$2 & 1.18E$-$2 & 2.05E$-$3 & 2.43E$-$4 \\
\hline
625 & 85 & 90 & 1 & 1 & 0.251 & 3.00 & 1.50 & 1.34 & 1.0117E$+$1 & 1.6189E$-$5 & 7.1502E$-$3 & 2048 & 10 & 64 & 40.0 & 9.15E$-$1 & 7.96E$-$2 & 1.36E$-$2 & 2.38E$-$5 \\
625 & 85 & 90 & 1 & 1 & 0.250 & 4.00 & 2.00 & 1.79 & 2.3774E$+$0 & 1.6189E$-$5 & 1.4749E$-$2 & 2048 & 10 & 64 & 40.0 & 4.91E$-$1 & 1.61E$-$1 & 1.70E$-$2 & 5.54E$-$5 \\
625 & 85 & 90 & 1 & 1 & 0.250 & 5.00 & 2.50 & 2.23 & 1.1237E$+$0 & 1.1332E$-$5 & 1.7949E$-$2 & 2048 & 10 & 64 & 25.2 & 2.79E$-$1 & 1.57E$-$1 & 1.36E$-$2 & 7.79E$-$5 \\
625 & 85 & 90 & 1 & 1 & 0.250 & 7.00 & 3.50 & 3.13 & 4.5493E$-$1 & 8.0944E$-$6 & 2.3840E$-$2 & 2048 & 10 & 64 & 20.0 & 1.35E$-$1 & 4.98E$-$2 & 1.35E$-$2 & 1.60E$-$4 \\
625 & 85 & 90 & 1 & 1 & 0.250 & 10.00 & 5.00 & 4.47 & 1.9912E$-$1 & 4.8566E$-$6 & 2.7912E$-$2 & 2048 & 10 & 64 & 15.2 & 1.29E$-$1 & 3.69E$-$2 & 1.75E$-$2 & 2.17E$-$4 \\
\hline
625 & 80 & 90 & 1 & 1 & 0.251 & 2.99 & 1.50 & 1.34 & 1.0117E$+$1 & 1.6189E$-$5 & 7.1502E$-$3 & 2048 & 10 & 64 & 30.1 & 8.73E$-$1 & 4.51E$-$2 & 1.27E$-$2 & 2.32E$-$5 \\
625 & 80 & 90 & 1 & 1 & 0.250 & 3.99 & 2.00 & 1.79 & 2.3774E$+$0 & 1.6189E$-$5 & 1.4749E$-$2 & 2048 & 10 & 64 & 30.1 & 6.58E$-$1 & 1.09E$-$1 & 1.94E$-$2 & 4.72E$-$5 \\
625 & 80 & 90 & 1 & 1 & 0.250 & 4.99 & 2.50 & 2.23 & 1.1237E$+$0 & 1.1332E$-$5 & 1.7949E$-$2 & 2048 & 10 & 64 & 25.2 & 5.28E$-$1 & 1.62E$-$1 & 2.13E$-$2 & 6.45E$-$5 \\
625 & 80 & 90 & 1 & 1 & 0.250 & 6.99 & 3.49 & 3.13 & 4.5493E$-$1 & 8.0944E$-$6 & 2.3840E$-$2 & 2048 & 10 & 64 & 20.0 & 2.38E$-$1 & 1.27E$-$1 & 2.05E$-$2 & 1.38E$-$4 \\
625 & 80 & 90 & 1 & 1 & 0.250 & 9.99 & 5.00 & 4.47 & 1.9912E$-$1 & 4.8566E$-$6 & 2.7912E$-$2 & 2048 & 10 & 64 & 15.2 & 1.89E$-$1 & 8.87E$-$2 & 2.27E$-$2 & 1.91E$-$4 \\
\hline
625 & 75 & 90 & 1 & 1 & 0.251 & 2.98 & 1.49 & 1.34 & 1.0117E$+$1 & 1.6189E$-$5 & 7.1502E$-$3 & 2048 & 10 & 64 & 40.0 & 8.54E$-$1 & 2.44E$-$2 & 1.27E$-$2 & 2.37E$-$5 \\
625 & 75 & 90 & 1 & 1 & 0.250 & 3.98 & 1.99 & 1.78 & 2.3774E$+$0 & 1.6189E$-$5 & 1.4749E$-$2 & 2048 & 10 & 64 & 30.1 & 6.45E$-$1 & 7.82E$-$2 & 1.94E$-$2 & 4.80E$-$5 \\
625 & 75 & 90 & 1 & 1 & 0.250 & 4.98 & 2.49 & 2.23 & 1.1237E$+$0 & 1.1332E$-$5 & 1.7949E$-$2 & 2048 & 10 & 64 & 25.2 & 7.76E$-$1 & 1.81E$-$1 & 2.69E$-$2 & 5.55E$-$5 \\
625 & 75 & 90 & 1 & 1 & 0.250 & 6.98 & 3.49 & 3.12 & 4.5493E$-$1 & 8.0944E$-$6 & 2.3840E$-$2 & 2048 & 10 & 64 & 20.0 & 1.11E$+$0 & 1.90E$-$1 & 5.52E$-$2 & 7.95E$-$5 \\
625 & 75 & 90 & 1 & 1 & 0.250 & 9.98 & 4.99 & 4.46 & 1.9912E$-$1 & 4.8566E$-$6 & 2.7912E$-$2 & 2048 & 10 & 64 & 15.2 & 2.58E$-$1 & 1.27E$-$1 & 3.09E$-$2 & 1.92E$-$4 \\
\hline
20 & 90 & 90 & 720 & 1 & 0.250 & 3.00 & 1.50 & 1.34 & 9.8755E$+$0 & 5.0590E$-$4 & 3.9505E$-$2 & 32 & 10 & 32 & 39.2 & 8.62E$-$1 & 9.55E$-$2 & 1.29E$-$2 & 7.46E$-$4 \\
20 & 90 & 90 & 720 & 1 & 0.250 & 4.00 & 2.00 & 1.79 & 2.2975E$+$0 & 5.0590E$-$4 & 8.1851E$-$2 & 32 & 10 & 32 & 39.2 & 4.43E$-$1 & 1.69E$-$1 & 1.58E$-$2 & 1.79E$-$3 \\
20 & 90 & 90 & 720 & 1 & 0.250 & 5.00 & 2.50 & 2.24 & 1.0872E$+$0 & 3.5413E$-$4 & 9.9512E$-$2 & 32 & 10 & 32 & 29.5 & 2.83E$-$1 & 1.40E$-$1 & 1.38E$-$2 & 2.44E$-$3 \\
20 & 90 & 90 & 720 & 1 & 0.250 & 6.84 & 3.42 & 3.06 & 4.6363E$-$1 & 2.5295E$-$4 & 1.2868E$-$1 & 32 & 10 & 32 & 24.5 & 2.13E$-$1 & 8.37E$-$2 & 1.73E$-$2 & 4.07E$-$3 \\
20 & 90 & 90 & 720 & 1 & 0.250 & 9.93 & 4.97 & 4.44 & 1.9289E$-$1 & 1.5177E$-$4 & 1.5439E$-$1 & 32 & 10 & 32 & 19.6 & 1.85E$-$1 & 8.66E$-$2 & 2.34E$-$2 & 6.33E$-$3 \\
20 & 90 & 90 & 720 & 1 & 0.250 & 15.12 & 7.56 & 6.76 & 7.9463E$-$2 & 5.0590E$-$5 & 1.3896E$-$1 & 32 & 10 & 32 & 14.7 & 1.70E$-$1 & 4.14E$-$2 & 1.82E$-$2 & 5.36E$-$3 \\
20 & 90 & 90 & 960 & 1 & 0.250 & 20.27 & 10.14 & 9.07 & 4.3483E$-$2 & 2.5295E$-$5 & 1.3285E$-$1 & 32 & 10 & 32 & 14.7 & 1.64E$-$1 & 4.36E$-$2 & 1.69E$-$2 & 5.15E$-$3 \\
\hline
49 & 90 & 90 & 1440 & 1 & 0.250 & 3.00 & 1.50 & 1.34 & 1.0020E$+$1 & 2.0649E$-$4 & 2.5420E$-$2 & 32 & 10 & 32 & 39.5 & 8.53E$-$1 & 1.09E$-$1 & 1.28E$-$2 & 3.07E$-$4 \\
49 & 90 & 90 & 1440 & 1 & 0.250 & 4.00 & 2.00 & 1.79 & 2.3453E$+$0 & 2.0649E$-$4 & 5.2528E$-$2 & 32 & 10 & 32 & 39.5 & 4.78E$-$1 & 2.11E$-$1 & 1.59E$-$2 & 6.78E$-$4 \\
49 & 90 & 90 & 1440 & 1 & 0.250 & 5.00 & 2.50 & 2.24 & 1.1090E$+$0 & 1.4454E$-$4 & 6.3899E$-$2 & 32 & 10 & 32 & 25.0 & 3.37E$-$1 & 1.71E$-$1 & 1.54E$-$2 & 9.32E$-$4 \\
49 & 90 & 90 & 1440 & 1 & 0.250 & 6.85 & 3.42 & 3.06 & 4.7256E$-$1 & 1.0325E$-$4 & 8.2703E$-$2 & 32 & 10 & 32 & 25.6 & 3.26E$-$1 & 1.43E$-$1 & 2.35E$-$2 & 1.47E$-$3 \\
49 & 90 & 90 & 1440 & 1 & 0.250 & 9.97 & 4.98 & 4.46 & 1.9696E$-$1 & 6.1947E$-$5 & 9.9192E$-$2 & 32 & 10 & 32 & 25.0 & 2.75E$-$1 & 6.97E$-$2 & 3.22E$-$2 & 2.39E$-$3 \\
49 & 90 & 90 & 720 & 1 & 0.250 & 15.16 & 7.58 & 6.78 & 8.0625E$-$2 & 2.0649E$-$5 & 8.9530E$-$2 & 128 & 10 & 32 & 15.0 & 1.81E$-$1 & 4.46E$-$2 & 1.92E$-$2 & 2.16E$-$3 \\
49 & 90 & 90 & 720 & 1 & 0.250 & 20.33 & 10.17 & 9.09 & 4.4032E$-$2 & 1.0325E$-$5 & 8.5672E$-$2 & 128 & 10 & 32 & 14.9 & 1.32E$-$1 & 3.76E$-$2 & 1.42E$-$2 & 2.19E$-$3 \\
49 & 90 & 90 & 360 & 1 & 0.250 & 30.63 & 15.32 & 13.70 & 1.9157E$-$2 & 4.1298E$-$6 & 8.2153E$-$2 & 128 & 10 & 32 & 10.0 & 1.14E$-$1 & 3.67E$-$2 & 1.11E$-$2 & 1.97E$-$3 \\
\hline
200 & 90 & 90 & 1440 & 1 & 0.250 & 3.00 & 1.50 & 1.34 & 1.0099E$+$1 & 5.0590E$-$5 & 1.2629E$-$2 & 32 & 10 & 32 & 33.1 & 8.54E$-$1 & 1.16E$-$1 & 1.28E$-$2 & 7.49E$-$5 \\
200 & 90 & 90 & 1440 & 1 & 0.250 & 4.00 & 2.00 & 1.79 & 2.3715E$+$0 & 5.0590E$-$5 & 2.6060E$-$2 & 32 & 10 & 32 & 39.3 & 5.40E$-$1 & 2.04E$-$1 & 1.80E$-$2 & 1.66E$-$4 \\
200 & 90 & 90 & 1440 & 1 & 0.250 & 5.00 & 2.50 & 2.24 & 1.1210E$+$0 & 3.5413E$-$5 & 3.1711E$-$2 & 32 & 10 & 32 & 29.0 & 4.75E$-$1 & 2.00E$-$1 & 2.03E$-$2 & 2.14E$-$4 \\
200 & 90 & 90 & 1440 & 1 & 0.250 & 6.86 & 3.43 & 3.07 & 4.7744E$-$1 & 2.5295E$-$5 & 4.1064E$-$2 & 32 & 10 & 32 & 21.5 & 3.66E$-$1 & 1.52E$-$1 & 2.67E$-$2 & 3.65E$-$4 \\
200 & 90 & 90 & 1440 & 1 & 0.250 & 9.98 & 4.99 & 4.46 & 1.9918E$-$1 & 1.5177E$-$5 & 4.9241E$-$2 & 32 & 10 & 32 & 20.0 & 3.73E$-$1 & 1.16E$-$1 & 3.91E$-$2 & 5.24E$-$4 \\
200 & 90 & 90 & 1440 & 1 & 0.250 & 15.18 & 7.59 & 6.79 & 8.1259E$-$2 & 5.0590E$-$6 & 4.4512E$-$2 & 32 & 10 & 32 & 15.0 & 2.44E$-$1 & 7.44E$-$2 & 2.51E$-$2 & 5.14E$-$4 \\
200 & 90 & 90 & 1440 & 1 & 0.250 & 20.36 & 10.18 & 9.11 & 4.4331E$-$2 & 2.5295E$-$6 & 4.2614E$-$2 & 64 & 10 & 32 & 10.2 & 1.51E$-$1 & 6.25E$-$2 & 1.55E$-$2 & 5.13E$-$4 \\
\hline
49 & 90 & 90 & 192 & 192 & 0.250 & 4.00 & 2.00 & 1.79 & 2.3453E$+$0 & 2.0649E$-$4 & 5.2528E$-$2 & 32 & 10 & 32 & 24.1 & 4.65E$-$1 & 2.23E$-$1 & 1.57E$-$2 & 6.90E$-$4 \\
49 & 90 & 90 & 192 & 192 & 0.250 & 5.00 & 2.50 & 2.24 & 1.1090E$+$0 & 1.4454E$-$4 & 6.3899E$-$2 & 32 & 10 & 32 & 35.3 & 3.37E$-$1 & 2.07E$-$1 & 1.57E$-$2 & 9.54E$-$4 \\
49 & 90 & 90 & 192 & 384 & 0.250 & 6.85 & 3.42 & 3.06 & 4.7256E$-$1 & 1.0325E$-$4 & 8.2703E$-$2 & 32 & 10 & 32 & 17.4 & 2.74E$-$1 & 1.40E$-$1 & 2.12E$-$2 & 1.58E$-$3 \\
49 & 90 & 90 & 192 & 192 & 0.250 & 9.97 & 4.98 & 4.46 & 1.9696E$-$1 & 6.1947E$-$5 & 9.9192E$-$2 & 32 & 10 & 32 & 15.0 & 2.37E$-$1 & 1.61E$-$1 & 2.84E$-$2 & 2.45E$-$3 \\
49 & 90 & 90 & 192 & 192 & 0.250 & 15.16 & 7.58 & 6.78 & 8.0625E$-$2 & 2.0649E$-$5 & 8.9530E$-$2 & 32 & 10 & 32 & 8.9 & 1.50E$-$1 & 7.62E$-$2 & 1.58E$-$2 & 2.16E$-$3 \\
\hline
49 & 90 & 90 & 1440 & 1 & 0.125 & 4.00 & 1.41 & 1.33 & 1.1537E$+$1 & 2.0649E$-$4 & 3.3500E$-$2 & 32 & 10 & 32 & 24.8 & 7.71E$-$1 & 1.32E$-$1 & 1.28E$-$2 & 3.38E$-$4 \\
49 & 90 & 90 & 1440 & 1 & 0.125 & 5.00 & 1.77 & 1.67 & 3.4565E$+$0 & 1.4454E$-$4 & 5.1197E$-$2 & 32 & 10 & 32 & 24.8 & 4.14E$-$1 & 1.98E$-$1 & 1.06E$-$2 & 5.22E$-$4 \\
49 & 90 & 90 & 1440 & 1 & 0.125 & 6.68 & 2.36 & 2.23 & 1.2431E$+$0 & 1.0325E$-$4 & 7.2128E$-$2 & 32 & 10 & 32 & 25.0 & 2.42E$-$1 & 1.35E$-$1 & 1.14E$-$2 & 9.66E$-$4 \\
49 & 90 & 90 & 1440 & 1 & 0.125 & 9.78 & 3.46 & 3.26 & 4.4927E$-$1 & 6.1947E$-$5 & 9.2894E$-$2 & 32 & 10 & 32 & 24.8 & 3.09E$-$1 & 1.34E$-$1 & 2.68E$-$2 & 1.77E$-$3 \\
49 & 90 & 90 & 1440 & 1 & 0.125 & 14.99 & 5.30 & 5.00 & 1.7122E$-$1 & 2.0649E$-$5 & 8.6888E$-$2 & 32 & 10 & 32 & 20.1 & 2.65E$-$1 & 8.51E$-$2 & 2.32E$-$2 & 1.79E$-$3 \\
49 & 90 & 90 & 1440 & 1 & 0.125 & 20.19 & 7.14 & 6.73 & 9.1148E$-$2 & 1.0325E$-$5 & 8.4213E$-$2 & 32 & 10 & 32 & 15.1 & 2.10E$-$1 & 6.01E$-$2 & 1.88E$-$2 & 1.83E$-$3 \\
49 & 90 & 90 & 360 & 1 & 0.125 & 30.53 & 10.79 & 10.18 & 3.8916E$-$2 & 4.1298E$-$6 & 8.1516E$-$2 & 128 & 10 & 32 & 10.1 & 1.47E$-$1 & 8.28E$-$2 & 1.36E$-$2 & 1.89E$-$3 \\
\hline
49 & 90 & 90 & 720 & 1 & 0.500 & 2.00 & 1.41 & 1.15 & 3.5805E$+$1 & 2.0649E$-$4 & 9.5092E$-$3 & 32 & 10 & 32 & 38.8 & 9.95E$-$1 & 6.62E$-$2 & 1.14E$-$2 & 2.34E$-$4 \\
49 & 90 & 90 & 1440 & 1 & 0.500 & 3.00 & 2.12 & 1.73 & 2.2310E$+$0 & 2.0649E$-$4 & 3.8088E$-$2 & 32 & 10 & 32 & 38.8 & 6.17E$-$1 & 1.70E$-$1 & 1.54E$-$2 & 5.09E$-$4 \\
49 & 90 & 90 & 1440 & 1 & 0.500 & 3.87 & 2.74 & 2.24 & 9.2639E$-$1 & 2.0649E$-$4 & 5.9093E$-$2 & 32 & 10 & 32 & 25.9 & 3.99E$-$1 & 1.41E$-$1 & 1.93E$-$2 & 9.85E$-$4 \\
49 & 90 & 90 & 1440 & 1 & 0.500 & 4.90 & 3.47 & 2.83 & 4.8236E$-$1 & 1.4454E$-$4 & 6.8506E$-$2 & 32 & 10 & 32 & 26.0 & 4.00E$-$1 & 1.29E$-$1 & 2.30E$-$2 & 1.17E$-$3 \\
49 & 90 & 90 & 1440 & 1 & 0.500 & 6.98 & 4.93 & 4.03 & 2.0639E$-$1 & 1.0325E$-$4 & 8.8477E$-$2 & 32 & 10 & 32 & 25.6 & 2.80E$-$1 & 5.22E$-$2 & 2.81E$-$2 & 2.04E$-$3 \\
49 & 90 & 90 & 1440 & 1 & 0.500 & 10.08 & 7.13 & 5.82 & 9.2084E$-$2 & 6.1947E$-$5 & 1.0257E$-$1 & 32 & 10 & 32 & 25.5 & 2.05E$-$1 & 3.55E$-$2 & 2.94E$-$2 & 2.94E$-$3 \\
49 & 90 & 90 & 1440 & 1 & 0.500 & 15.26 & 10.79 & 8.81 & 3.9095E$-$2 & 2.0649E$-$5 & 9.0910E$-$2 & 32 & 10 & 32 & 19.9 & 1.47E$-$1 & 1.99E$-$2 & 1.78E$-$2 & 2.47E$-$3 \\
\hline
49 & 90 & 90 & 1440 & 1 & 1.000 & 2.00 & 2.00 & 1.41 & 4.4802E$+$0 & 2.0649E$-$4 & 1.9008E$-$2 & 32 & 10 & 32 & 25.0 & 9.19E$-$1 & 1.39E$-$1 & 1.34E$-$2 & 2.97E$-$4 \\
49 & 90 & 90 & 1440 & 1 & 1.000 & 3.00 & 3.00 & 2.12 & 8.2034E$-$1 & 2.0649E$-$4 & 4.4412E$-$2 & 32 & 10 & 32 & 25.0 & 5.18E$-$1 & 1.55E$-$1 & 1.76E$-$2 & 6.95E$-$4 \\
49 & 90 & 90 & 1440 & 1 & 1.000 & 4.00 & 4.00 & 2.83 & 3.6299E$-$1 & 2.0649E$-$4 & 6.6745E$-$2 & 32 & 10 & 32 & 25.0 & 3.64E$-$1 & 1.08E$-$1 & 2.37E$-$2 & 1.33E$-$3 \\
49 & 90 & 90 & 1440 & 1 & 1.000 & 4.99 & 4.99 & 3.53 & 2.1083E$-$1 & 1.4454E$-$4 & 7.3265E$-$2 & 32 & 10 & 32 & 24.8 & 3.42E$-$1 & 8.28E$-$2 & 2.55E$-$2 & 1.52E$-$3 \\
49 & 90 & 90 & 1440 & 1 & 1.000 & 7.06 & 7.06 & 4.99 & 9.6333E$-$2 & 1.0325E$-$4 & 9.1569E$-$2 & 32 & 10 & 32 & 24.8 & 2.18E$-$1 & 3.84E$-$2 & 2.54E$-$2 & 2.37E$-$3 \\
49 & 90 & 90 & 1440 & 1 & 1.000 & 10.15 & 10.15 & 7.18 & 4.4490E$-$2 & 6.1947E$-$5 & 1.0434E$-$1 & 32 & 10 & 32 & 25.0 & 1.59E$-$1 & 2.53E$-$2 & 2.49E$-$2 & 3.19E$-$3 \\
49 & 90 & 90 & 1440 & 1 & 1.000 & 15.31 & 15.31 & 10.82 & 1.9246E$-$2 & 2.0649E$-$5 & 9.1619E$-$2 & 32 & 10 & 32 & 24.8 & 1.32E$-$1 & 1.91E$-$2 & 1.64E$-$2 & 2.53E$-$3 \\
\hline
49 & 90 & 90 & 1440 & 1 & 2.000 & 2.00 & 2.83 & 1.63 & 1.4344E$+$0 & 2.0649E$-$4 & 2.3754E$-$2 & 32 & 10 & 32 & 24.8 & 8.56E$-$1 & 1.86E$-$1 & 1.50E$-$2 & 3.58E$-$4 \\
49 & 90 & 90 & 1440 & 1 & 2.000 & 3.00 & 4.24 & 2.45 & 3.5747E$-$1 & 2.0649E$-$4 & 4.7572E$-$2 & 32 & 10 & 32 & 24.8 & 5.24E$-$1 & 9.73E$-$2 & 2.04E$-$2 & 7.93E$-$4 \\
49 & 90 & 90 & 1512 & 1 & 2.000 & 4.00 & 5.66 & 3.27 & 1.6927E$-$1 & 2.0649E$-$4 & 6.9110E$-$2 & 32 & 10 & 32 & 27.8 & 3.39E$-$1 & 4.34E$-$2 & 2.63E$-$2 & 1.59E$-$3 \\
49 & 90 & 90 & 1512 & 1 & 2.000 & 5.00 & 7.07 & 4.08 & 1.0012E$-$1 & 2.0649E$-$4 & 8.9824E$-$2 & 32 & 10 & 32 & 24.6 & 2.58E$-$1 & 3.86E$-$2 & 3.21E$-$2 & 2.55E$-$3 \\
49 & 90 & 90 & 1440 & 1 & 2.000 & 7.11 & 10.05 & 5.80 & 4.6506E$-$2 & 1.0325E$-$4 & 9.3186E$-$2 & 32 & 10 & 32 & 24.9 & 1.86E$-$1 & 2.32E$-$2 & 2.42E$-$2 & 2.66E$-$3 \\
49 & 90 & 90 & 2016 & 1 & 2.000 & 10.19 & 14.41 & 8.32 & 2.1861E$-$2 & 6.1947E$-$5 & 1.0525E$-$1 & 32 & 10 & 32 & 18.8 & 1.33E$-$1 & 1.29E$-$2 & 2.26E$-$2 & 3.46E$-$3 \\
49 & 90 & 90 & 1440 & 1 & 2.000 & 15.33 & 21.69 & 12.52 & 9.5477E$-$3 & 2.0649E$-$5 & 9.1978E$-$2 & 32 & 10 & 32 & 12.0 & 1.22E$-$1 & 2.36E$-$2 & 1.57E$-$2 & 2.63E$-$3 \\
\hline
49 & 90 & 90 & 1440 & 1 & 4.000 & 2.00 & 4.00 & 1.79 & 5.9285E$-$1 & 2.0649E$-$4 & 2.6126E$-$2 & 32 & 10 & 32 & 25.1 & 8.35E$-$1 & 1.87E$-$1 & 1.62E$-$2 & 3.96E$-$4 \\
49 & 90 & 90 & 1512 & 1 & 4.000 & 3.00 & 6.00 & 2.68 & 1.6743E$-$1 & 2.0649E$-$4 & 4.9151E$-$2 & 32 & 10 & 32 & 23.8 & 4.90E$-$1 & 6.76E$-$2 & 2.21E$-$2 & 9.19E$-$4 \\
49 & 90 & 90 & 1512 & 1 & 4.000 & 4.00 & 8.00 & 3.58 & 8.1810E$-$2 & 2.0649E$-$4 & 7.0292E$-$2 & 32 & 10 & 32 & 34.3 & 3.45E$-$1 & 4.17E$-$2 & 2.79E$-$2 & 1.65E$-$3 \\
49 & 90 & 90 & 1512 & 1 & 4.000 & 5.00 & 10.00 & 4.47 & 4.9023E$-$2 & 2.0649E$-$4 & 9.0767E$-$2 & 32 & 10 & 32 & 36.6 & 2.60E$-$1 & 2.30E$-$2 & 3.32E$-$2 & 2.61E$-$3 \\
49 & 90 & 90 & 1440 & 1 & 4.000 & 7.13 & 14.27 & 6.38 & 2.2843E$-$2 & 1.0325E$-$4 & 9.4017E$-$2 & 32 & 10 & 32 & 25.1 & 1.69E$-$1 & 1.46E$-$2 & 2.31E$-$2 & 2.79E$-$3 \\
49 & 90 & 90 & 2160 & 1 & 4.000 & 10.21 & 20.42 & 9.13 & 1.0835E$-$2 & 6.1947E$-$5 & 1.0571E$-$1 & 32 & 10 & 32 & 19.0 & 1.35E$-$1 & 1.22E$-$2 & 2.33E$-$2 & 3.52E$-$3 \\

\enddata
\tablecomments{
Table~\ref{tab:param} is
available in a machine-readable CSV format in the online journal.
}
\end{deluxetable*}
\end{longrotatetable}

\end{document}